\documentclass[aps,prxquantum,twocolumn,superscriptaddress,nofootinbib]{revtex4-2}
\usepackage[table, svgnames, dvipsnames]{xcolor}
\usepackage{makecell, cellspace}
\usepackage{bbold}
\usepackage{amsfonts}

\usepackage{bbm}
\usepackage{mathdesign}
\usepackage{mathtools}
\usepackage{amsthm}
\usepackage{bm}
\usepackage{amssymb}
\usepackage[normalem]{ulem}
\usepackage{braket}
\usepackage{hyperref}
\usepackage{tikz}
\usepackage{ifthen}
\usepackage{stmaryrd}
\usetikzlibrary{tikzmark}
\usepackage{upgreek}
\usepackage{physics}

\usepackage[]{lineno}

\usepackage{svg}

\usepackage{booktabs, siunitx}

\usepackage[
    top=3cm,    
    bottom=3cm, 
    left=2cm, 
    right=2cm 
]{geometry}
\usepackage{soul}

\usepackage{ragged2e} 

\usepackage{graphicx}
\usepackage[shortlabels]{enumitem}

\usepackage{tikz-cd}
\usepackage{systeme}
\usepackage{tensor}
\usepackage{xparse}
\usepackage{amsmath}
\usepackage{mathtools}
\usepackage{amssymb}
\usepackage{dsfont}
\usepackage[table]{xcolor}
\usepackage{upgreek}
\usepackage{mathrsfs}
\usetikzlibrary{decorations.pathreplacing}
\definecolor{mycolor}{HTML}{90E0EF}

\usepackage{lettrine}
\usepackage{setspace}
\usepackage[utf8]{inputenc} 
\usepackage[T1]{fontenc}

\usepackage{longtable}

\usepackage{wrapfig}

\theoremstyle{plain}

\theoremstyle{remark}

\renewcommand{\selectlanguage}[1]{}

\definecolor{blue}{HTML}{03045E}
\definecolor{red}{HTML}{c1121f}

\definecolor{gblue}{HTML}{4285F4}

\usepackage{orcidlink}
\hypersetup{
	colorlinks=true,
	citecolor= blue,
	linkcolor=blue,
	filecolor=blue,      
	urlcolor=blue,
	pdftitle={Overleaf Example},
	pdfpagemode=FullScreen,
}

\usepackage[export]{adjustbox}

\usepackage{titlesec}
\titleformat{\section}
{\sffamily\bfseries} 
{\thesection} 
{1em} 
{} 

\titleformat{\subsection}
{\sffamily\bfseries} 
{\thesubsection} 
{1em} 
{} 

\titleformat{\subsubsection}
  {\normalsize\itshape\centering} 
  {\thesubsubsection.} 
  {1em} 
  {} 

\newcommand{\figref}[1]{\hyperref[#1]{Fig.~\ref{#1}}}
\newcommand{\figaref}[1]{\hyperref[#1]{Fig.~\ref{#1}(a)}}
\newcommand{\figbref}[1]{\hyperref[#1]{Fig.~\ref{#1}(b)}}
\newcommand{\figcref}[1]{\hyperref[#1]{Fig.~\ref{#1}(c)}}
\newcommand{\figdref}[1]{\hyperref[#1]{Fig.~\ref{#1}(d)}}

\titlespacing\section{0pt}{12pt plus 3pt minus 2pt}{5pt plus 2pt minus 2pt}
\titlespacing\subsection{0pt}{12pt plus 3pt minus 2pt}{5pt plus 2pt minus 2pt}
\titlespacing\subsubsection{0pt}{12pt plus 3pt minus 2pt}{5pt plus 2pt minus 2pt}
\usepackage{lipsum} 

\makeatletter
\let\@afterindenttrue\@afterindentfalse
\makeatother

\definecolor{ggreen}{HTML}{34A853}

\begin{document}
\title{\textbf{\textsf{
Graph-State Circuit Blocks control Entanglement and Scrambling Velocities}}}
\author{Chandana Rao$^{\orcidlink{0009-0000-4397-353X}}$}
\affiliation{Department of Computer Science, Paderborn University, Warburger Str.~100, 33098, Paderborn, Germany}
\affiliation{Institute for Photonic Quantum Systems (PhoQS), Paderborn University, Warburger Str.~100, 33098 Paderborn, Germany}

\author{Himanshu Sahu$^{\orcidlink{0000-0002-9522-6592}}$}
\email{hsahu@perimeterinstitute.ca}
\affiliation{Perimeter Institute for Theoretical Physics, Waterloo, ON, N2L 2Y5, Canada.}
\affiliation{Department of Physics and Astronomy and Institute for Quantum Computing, University of Waterloo, ON N2L 3G1, Canada.}

\author{Aranya Bhattacharya$^{\orcidlink{0000-0002-1882-4177}}$}
\affiliation{School of Mathematics, University of Bristol, Fry Building, Woodland Road, Bristol BS8 1UG, UK}
\affiliation{Institute of Theoretical Physics,  Jagiellonian University, Łojasiewicza 11, 30-348 Kraków, Poland}

\author{Suhail Ahmad Rather
}
\affiliation{Dahlem Center for Complex Quantum Systems, Freie Universit\"at Berlin, 14195 Berlin, Germany}

\author{Mario Flory$^{\orcidlink{0000-0002-7155-855X}}$}
\affiliation{Institute of Theoretical Physics,  Jagiellonian University, Łojasiewicza 11, 30-348 Kraków, Poland}

\author{Zahra Raissi$^{\orcidlink{0000-0002-9168-8212}}$}
\affiliation{Department for Quantum Technology, University of M\"unster, 48149 M\"unster, Germany}
\affiliation{Department of Computer Science, Paderborn University, Warburger Str.~100, 33098, Paderborn, Germany}
\affiliation{Institute for Photonic Quantum Systems (PhoQS), Paderborn University, Warburger Str.~100, 33098 Paderborn, Germany}

\begin{abstract}
\textbf{\textsf{Abstract}}: Random circuit models often describe local dynamics using generic two-qubit gates, which have proven successful in capturing entanglement growth and operator spreading in many contexts. This approach naturally leads to the expectation that detailed gate structure plays only a limited role in coarse-grained entanglement and scrambling diagnostics. We show that the internal structure of multipartite circuit primitives can significantly influence these dynamical rates, even within a fixed random-circuit architecture.
To investigate this, we study an exactly simulable family of Clifford quantum circuits built from fixed $n$-qubit graph-state preparation unitaries, which we treat as elementary building blocks. Specifically, we consider a one-dimensional chain of $N$ qubits initialized in a product state and evolved by layers in which nonoverlapping length-$n$ blocks are placed at uniformly random positions with sparsity $\alpha$. We find that different choices of graph-state building blocks lead to strongly varying dynamical rates. Graph states that are inequivalent under local Clifford (LC) transformations generate sharply different entanglement velocities $v_E$ (defined from the growth rate of bipartite entanglement) and butterfly velocities $v_B$ (defined from the propagation of out-of-time-order correlators), even though the circuits are drawn from the same ensemble with identical architecture and randomness parameters.
We further show that this hierarchy is captured by two complementary block-level characteristics: the distribution of entanglement across internal bipartitions of the graph state, which correlates with $v_E$, and a graph-theoretic connectivity profile across bipartitions, which correlates with $v_B$. Neither descriptor alone fully determines the dynamics; rather, entanglement growth and operator spreading are controlled by distinct structural features of the local circuit blocks. Notably, Absolutely Maximally Entangled (AME) states appear among the fastest scrambling building blocks within the ensembles studied here.

\vspace{2em}

\end{abstract}

\maketitle

\begin{figure*}[hbt!]
    \centering
\includegraphics[width=0.98\linewidth]{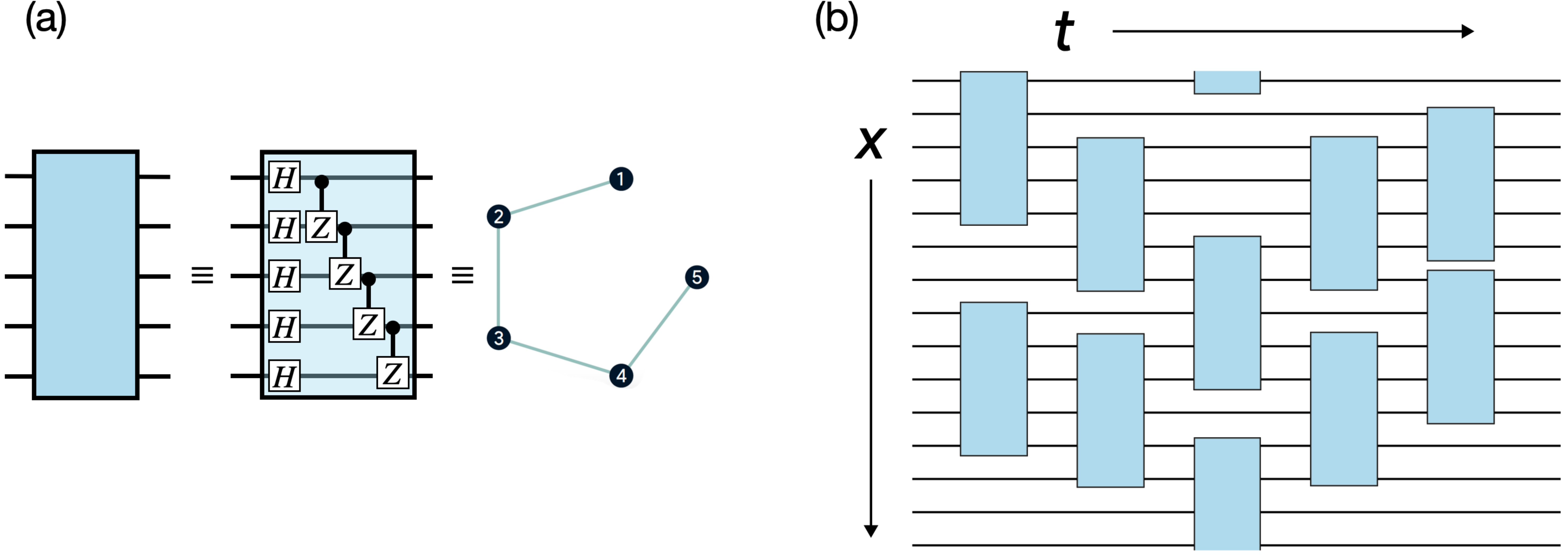}
\caption{
Graph-state circuit blocks used as local building blocks of the model. 
(a) Each blue block represents a preparation unitary for an $n$-qubit graph state $|\mathcal{G}\rangle$, starting from $\ket{0}^{\otimes n}$. In the graph representation, vertices correspond to qubits and edges indicate controlled-$Z$ entangling connections, specifying the internal entanglement structure of the block. Different choices of graph states that are not related by local Clifford (LC) transformations define distinct circuit models. 
(b) Schematic of the random Clifford circuits constructed from these blocks: the same $n$-qubit building block is applied at randomly chosen, nonoverlapping positions along a one-dimensional qubit chain of size $N$.
The density of applied blocks in each layer is controlled by the sparsity parameter $\alpha$.}   \label{fig:block-structure}
\end{figure*}

\section{Introduction}\label{sec:introduction}

Understanding how quantum information spreads and becomes scrambled under local unitary dynamics is a central problem in many-body quantum physics~\cite{Xu-Swingle-PRXQ-2024,fisher_random_2023, miInformationScramblingQuantum2021,Swingle2018Unscrambling,Nahum-PhysRevX-2018,25v1-my1x}.
A widely used approach is to study random quantum circuits, which provide minimal models of generic interacting dynamics in which unitary gates acting on neighboring qubits are drawn randomly from a sufficiently rich ensemble (e.g., Haar random, 2-designs)~\cite{Harrow-Low-2009,Brandao-Harrow-Horodecki-2016,PhysRevLett.132.250402,PhysRevB.102.064305,Nahum-PhysRevX-2017}.
Random circuit models have been shown to capture several robust and universal features of quantum chaos. In particular, the bipartite entanglement entropy of an initially weakly entangled state typically grows linearly in time before saturating to a volume-law value~\cite{Nahum-PhysRevX-2017,Zhou-Nahum-2019,Haferkamp_2022,RevModPhys.82.277}.
At the same time, local operators spread through the system with a finite characteristic velocity~\cite{liebFiniteGroupVelocity1972, guLocalCriticalityDiffusion2017, aleinerMicroscopicModelQuantum2016}, giving rise to an emergent light-cone structure that is commonly probed using out-of-time-ordered correlators (OTOCs)~\cite{Larkin1969,shenkerBlackHolesButterfly2014,maldacena2016bound,Xu-Swingle-PRXQ-2024,Toga2026AnalogOTOC}. The associated butterfly velocity $v_B$  characterizes the speed at which operator support and locally accessible quantum information propagate,~\cite{Xu-Swingle-PRXQ-2024,Nahum-PhysRevX-2018,Swingle2018Unscrambling}.
Together with the universal features of entanglement growth, these results motivate the common working assumption that coarse-grained dynamical diagnostics—such as entanglement and operator-spreading velocities—are largely insensitive to microscopic details of the local gates, and instead depend primarily on geometry and locality~\cite{Nahum-PhysRevX-2017,Nahum-PhysRevX-2018}.

Building on the random-circuit perspective, closely related questions have been explored in the context of quantum circuit complexity and the emergence of effective randomness in many-body dynamics. Motivated in part by black-hole thought experiments~\cite{patrickhaydenBlackHolesMirrors2007,shenkerBlackHolesButterfly2014}, it has been argued that local quantum circuits drawn from random gate ensembles can exhibit an extended regime of linear growth of circuit complexity before eventual saturation near its maximal value~\cite{Brown-Susskind-PhysRevD2018,Brandao-modelsofcomplexity-2021,Haferkamp_2022,25v1-my1x}.
In parallel, a precise mathematical framework has been developed to characterize how local circuit dynamics generate effective randomness. In particular, results on unitary designs show that local random circuits reproduce low-order statistical properties of Haar-random unitaries with increasing depth, despite the locality of the underlying gates~\cite{Harrow-Low-2009,Brandao-Harrow-Horodecki-2016}. Importantly, recent work~\cite{Brandao-Harrow-Horodecki-2016,suzuki-2025-moreglobalrandomness,Riddell:2025hbb} has demonstrated that such pseudorandom behavior does not require fully Haar-random local gates: ensembles with additional algebraic or structural constraints can generate comparable design properties at the level of low-order moments.
Taken together, these results indicate that strong pseudorandom behavior can emerge even in circuit models built from structured local primitives, suggesting that coarse-grained dynamical quantities—such as entanglement growth and operator spreading—may be governed primarily by architectural features rather than by microscopic gate details.

Recent work on random-circuit models has clarified how coarse dynamical quantities such as entanglement growth and operator spreading emerge from local unitary dynamics. In most existing studies~\cite{Nahum-PhysRevX-2017,Nahum-PhysRevX-2018,Zhou-Nahum-2019,Xu-Swingle-PRXQ-2024}, the elementary circuit update is taken to be a two-qubit gate drawn from a generic ensemble, with emphasis placed on how geometry and locality constrain quantities such as the entanglement velocity $v_E$ and the butterfly velocity $v_B$.
However, many experimentally relevant platforms naturally provide access to structured multipartite entangled resources rather than arbitrary Haar-random gates~\cite{PRXQuantum-Raissi,Economou_2010-2DCluster,Russo_2019-stategeneration,Coste_2023-stategeneration,Gimeno_Segovia_2019-stategeneration,Nguyen_2023-Rydberg,kim2022rydberg,Raussendorf-oneway}. Graph states provide a particularly well-studied and versatile class of such resources: they are stabilizer states with a simple graph-theoretic description~\cite{hein2006graphstatesapplications,hein-graphstates,Raissi-Barnes-2022,Raissi_Teixido-2020}, play a key role in measurement-based quantum computation~\cite{Raussendorf-oneway} and quantum error correction~\cite{gottesmanthesis,gottesman2009introductionQEC,Raissi-Gogolin-Riera-Acin-2018,Raissi-modified-2020}, and can be generated deterministically in a variety of physical systems.
 Motivated by this perspective, we study a class of local Clifford circuits in which the elementary operation is a fixed $n$-qubit graph-state preparation unitary, applied at random nonoverlapping positions along a one-dimensional qubit chain with tunable sparsity, see Fig.~\ref{fig:block-structure},  which illustrates both the internal structure of the graph-state blocks and their placement within the random circuit. This construction retains locality and exact classical simulability while allowing systematic control over the internal entanglement structure of the circuit’s building blocks~\cite{Aaronson-Gottesman,Van-den-Nest-Dehaene-Moor}.
 
Within this framework, we compare graph-state blocks that are inequivalent under local Clifford transformations and examine how their internal structure influences long-time dynamics. We find that different graph blocks give rise to distinct entanglement velocities $v_E$ and butterfly velocities $v_B$, even when the global circuit architecture is held fixed. These differences organize the blocks into a clear hierarchy of entanglement growth and operator spreading rates.  
In particular, within the graph-state blocks studied here, the blocks realizing absolutely maximally entangled (AME) states ~\cite{Raissi-Barnes-2022,Goyeneche_2018,AME-review,Raissi-Karimipour-2017,Raissi_Teixido-2020} exhibit the largest entanglement velocities $v_E$ in the cases where such states exist. At the same time, other graph structures can yield larger butterfly velocities $v_B$, depending on their connectivity structure.
 The detailed comparison across graph structures is discussed in Sec.~\ref{sec:results}.
Concretely, we (i) define a random circuit ensemble built from a fixed $n$-qubit graph-state preparation unitary placed at uniformly random, nonoverlapping locations with tunable sparsity $\alpha$; (ii) extract the entanglement and butterfly velocities $v_E$ and $v_B$ for system size  $N=200$ across multiple LC-inequivalent  graph-state blocks for block sizes $n=4,5,6$ (with additional results for $n=7$ reported in Appendix~\ref{appendix:results for seven-qubit graph}); and (iii) introduce two complementary block descriptors—a cut-entanglement profile ($\gamma$) and a cut-edge connectivity profile ($\wp$)—that correlate with $v_E$ and $v_B$, respectively, and clarify the regimes in which each descriptor provides predictive power.

To characterize the resulting dynamics, we focus on two complementary sets of quantities. On the dynamical side, we extract the entanglement velocity $v_E$ from the linear growth of bipartite entanglement and the butterfly velocity $v_B$ from the propagation of OTOCs. 
On the structural side, we introduce block-level descriptors that quantify (i) how entanglement is distributed across internal bipartitions of the graph-state block, captured by its cut-entanglement profile, and (ii) how efficiently operators can propagate across the block, captured by a graph-theoretic connectivity measure. We find that neither descriptor alone fully determines the dynamics; instead, entanglement growth and scrambling are jointly constrained by both internal entanglement structure and operator transport channels.

The remainder of the paper is organized as follows.
In Sec.~\ref{sec:background} we review graph states and the diagnostics used to characterize entanglement growth and operator spreading. 
In Sec.~\ref{sec:Model}, we introduce the random circuit ensemble built from graph-state blocks and define its key parameters. 
In Sec.~\ref{sec:diagnostics}, we define the entanglement and OTOC diagnostics and describe the procedures used to extract entanglement and butterfly velocities. In Sec.~\ref{sec:results}, we present numerical results demonstrating the hierarchy of entanglement and scrambling rates across locally inequivalent graph-state blocks and analyze their dependence on block structure. Finally, in Sec.~\ref{sec:discussion} we discuss the broader implications of the results and outline possible directions for future work.

\section{Background}\label{sec:background}

In this section we review the background required for the circuit construction and diagnostics used in this paper. We focus on graph states and their local Clifford structure, standard features of entanglement growth in random circuits, and operator spreading as quantified by out-of-time-ordered correlators. This presentation is intentionally concise and tailored to the quantities analyzed in later sections.

\subsection{Graph states and local Clifford structure}\label{subsec:graph_states}

Graph states form a central class of multipartite entangled stabilizer states with a simple graph-theoretic representation. Given a graph $\mathcal{G}=(\mathcal{V},\mathcal{E})$ with vertices $\mathcal{V}$ and edges $\mathcal{E}$, the corresponding graph state is defined as
\begin{equation} \label{eq:graph state}
    |\mathcal{G}\rangle = \prod_{(u,v)\in\mathcal{E}} \mathrm{CZ}_{uv}\, |+\rangle^{\otimes |\mathcal{V}|},
\end{equation}
where $\mathrm{CZ}_{uv}$ denotes the controlled-$Z$ gate acting on vertices $u$ and $v$. Equivalently, the state $|\mathcal{G}\rangle$ is the unique simultaneous $+1$ eigenstate of a set of commuting stabilizer generators
\begin{equation}
    S_u = X_u \prod_{v\in\mathcal{N}_u} Z_v ,
\end{equation}
where $\mathcal{N}_u$ denotes the neighborhood of vertex $u$ in $\mathcal{G}$~\cite{hein2006graphstatesapplications,Van-den-Nest-Dehaene-Moor}.

\begin{figure}[t]
    \centering
    \includegraphics[width=0.85\linewidth]{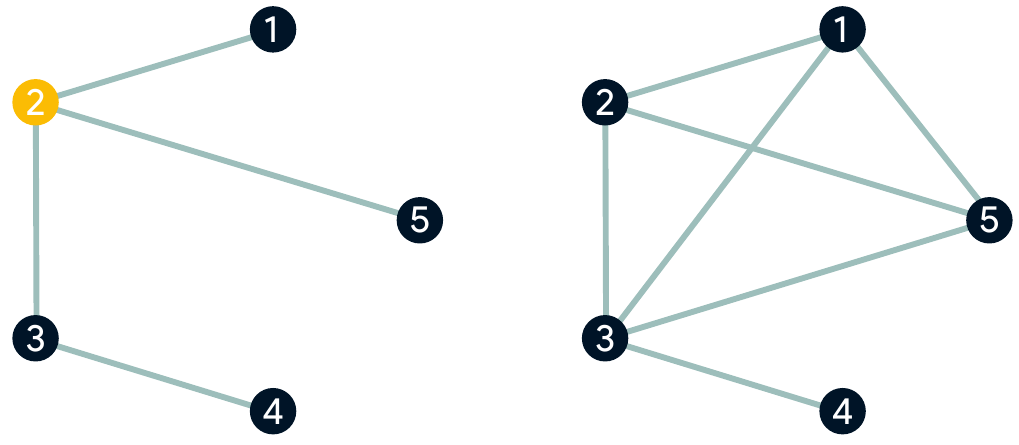}
    \caption{Illustration of local complementation on a five-qubit graph state. Local complementation at vertex $2$ (highlighted) transforms the graph on the left into the graph on the right by toggling the presence or absence of edges between all neighbors of that vertex. The two graphs are related by a local Clifford transformation.}
    \label{fig:local complementation}
\end{figure}

Two graph states are said to be local-Clifford (LC) equivalent if they are related by a product of single-qubit Clifford unitaries~\cite{hein2006graphstatesapplications}. LC equivalence is a restricted form of local unitary (LU) equivalence. For stabilizer states, LC equivalence always implies LU equivalence. Conversely, it has been shown that LU equivalence implies LC equivalence for stabilizer states on up to 26 qubits~\cite{VanDenNest2005,Ji:2008wfk,Claudet:2026mvt}. In this work, we classify graph-state blocks according to their LC-equivalence classes, which provide a tractable notion of structural equivalence for stabilizer circuit primitives.

\begin{figure}
    \centering
    \includegraphics[width=0.85\linewidth]{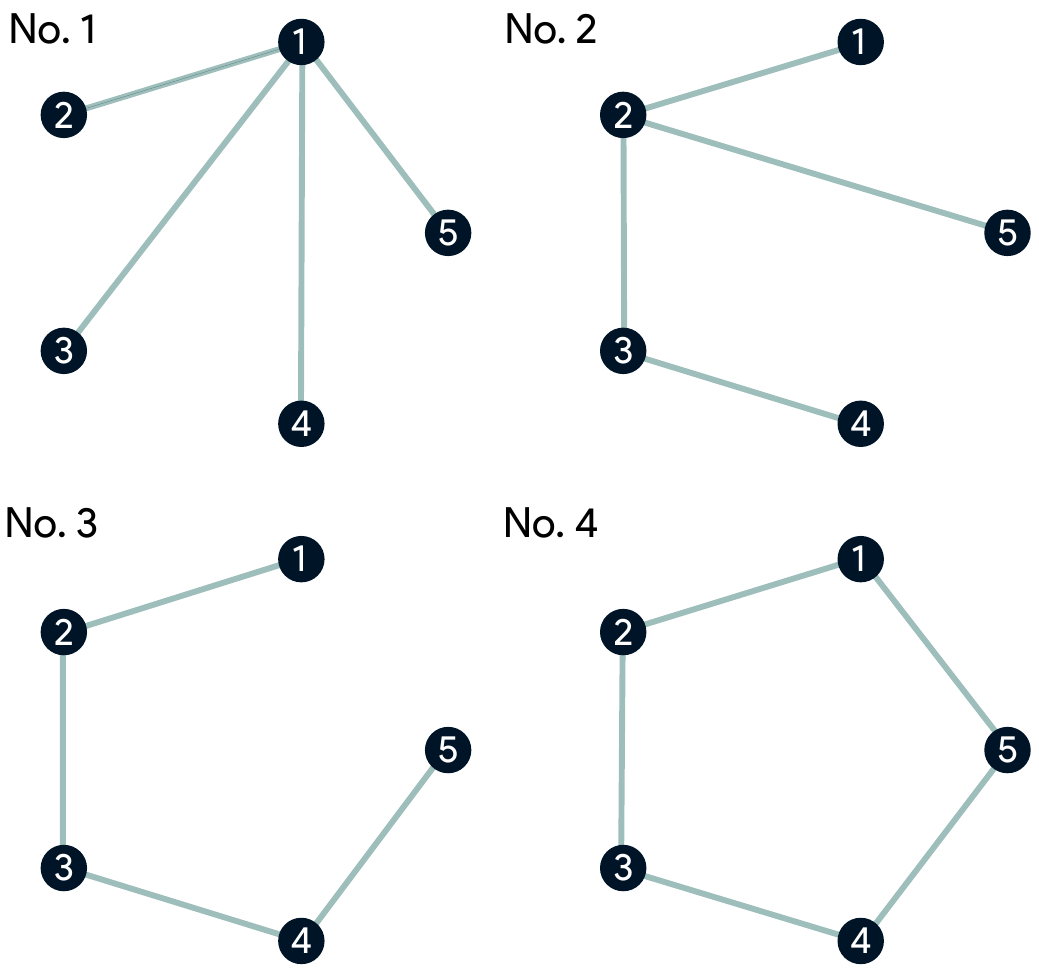}
    \caption{LC–inequivalent graph states for five qubits, shown as an illustrative example.
    Any other five-qubit graph state is LC-equivalent to one of these representatives.}
    \label{fig:5q-graph}
\end{figure}

A fundamental operation generating LC equivalence classes is local complementation: acting on a vertex $u$ toggles the presence or absence of edges between all pairs of neighbors of $u$, an explicit example is shown in Fig.~\ref{fig:local complementation}.
Repeated application of local complementation generates the full LC-equivalence class of a graph~\cite{hein2006graphstatesapplications,Van-den-Nest-Dehaene-Moor}. Throughout this paper, graph-state blocks that are not related by LC transformations are treated as structurally distinct resources.
LC transformations correspond to local Clifford basis changes, which preserve stabilizer entanglement properties while providing an operational notion of equivalence for graph-state circuit primitives.
An explicit example of LC–inequivalent five-qubit graph states is shown in Fig.~\ref{fig:5q-graph}.

Graph states play a prominent role in measurement-based quantum computation, quantum error correction, and multipartite entanglement theory. 
Of particular relevance here are absolutely maximally entangled (AME) states, which are multipartite states whose reduced density matrices on any subset of up to half the parties are maximally mixed when such states exist~\cite{Scott2004,AME-review,Goyeneche_2018,Raissi-Barnes-2022,Raissi-Karimipour-2017}. When exact AME states are not available for a given system size, one can still consider highly entangled graph states whose reduced density matrices on many subsystems are maximally mixed~\cite{Scott2004,Raissi_Teixido-2020}.

\subsection{Entanglement growth in random circuits}
\label{subsec:entanglement_background}

The generation of entanglement under local quantum dynamics is a defining feature of thermalization and scrambling in many-body systems~\cite{Amico-2008-many-body,RevModPhys.82.277}. For a pure state $|\Psi\rangle$ on a bipartition $A\cup\bar{A}$, entanglement is quantified by the von Neumann entropy
\begin{equation}
    S_A = -\mathrm{Tr}(\rho_A \ln \rho_A), \qquad \rho_A=\mathrm{Tr}_{\bar{A}} |\Psi\rangle\langle\Psi|.
\end{equation}

Random circuit models provide a minimal theoretical setting in which entanglement dynamics can be studied analytically and numerically~\cite{Nahum-PhysRevX-2017,Nahum-PhysRevX-2018,vonKeyserlingk2018operator,Zhou-Nahum-2019}. In one-dimensional circuits with local gates, the entanglement entropy of an initially weakly entangled state typically grows linearly in time before saturating to a volume-law value proportional to the subsystem size~\cite{Nahum-PhysRevX-2017,Zhou-Nahum-2019,Haferkamp_2022,vonKeyserlingk2018operator}. The coefficient of the linear growth defines the entanglement velocity $v_E$~\cite{Nahum-PhysRevX-2017,Nahum-PhysRevX-2018}.

This behavior admits an effective geometric interpretation in terms of minimal cuts or membranes in the spacetime tensor-network representation of the circuit~\cite{Nahum-PhysRevX-2017,Zhou-Nahum-2019}. In this picture, entanglement growth is governed by the cost of separating a region from its complement through the circuit geometry, providing an intuitive link between local gate structure and global entanglement dynamics. While this framework predicts robust linear growth, the value of $v_E$ can depend sensitively on the properties of the local gates, particularly when the gate ensemble deviates from fully generic randomness~\cite{Haferkamp_2022,suzuki-2025-moreglobalrandomness,PhysRevLett.132.250402}.

\subsection{Operator spreading and out-of-time-ordered correlators}
\label{subsec:otoc_background}

Complementary to entanglement growth, operator spreading characterizes how initially local information propagates through a quantum system~\cite{liebFiniteGroupVelocity1972,Swingle2018Unscrambling}. A standard diagnostic is the OTOC, which probes the noncommutativity of operators under time evolution. For local operators $W$ and $V_x$, the OTOC is defined as~\cite{Larkin1969,Nahum-PhysRevX-2018,maldacena2016bound}
\begin{equation}\label{eq:OTOC}
    C(x,t) = -\frac{1}{2}\,\overline{\mathrm{Tr}\big([W(t),V_x]^2\big)},
\end{equation}
where $W(t)=U^\dagger(t) W U(t)$, the overline denotes an average over circuit realizations or initial states and $V_x$ acts on the qubit located at position $x$.

In local quantum circuits, OTOCs typically exhibit a light-cone structure: outside a linearly expanding region they remain small, while inside the light cone they rapidly approach their maximal value~\cite{Nahum-PhysRevX-2018,Xu-Swingle-PRXQ-2024,vonKeyserlingk2018operator,Zhou-Nahum-2019}. The velocity of this front defines the butterfly velocity $v_B$, which characterizes the speed at which operator support and locally accessible quantum information propagate.

In Clifford circuits, operator dynamics simplify because Pauli operators remain Pauli operators under time evolution. 
Concretely, for any Pauli operator $W$, conjugation by a Clifford unitary $U$ yields
\begin{equation}
    U^\dagger W U = \pm P,
\end{equation}
where $P$ is another Pauli string. 
As a result, operator spreading can be tracked exactly by monitoring the spatial support of the evolved Pauli string, providing a characterization of operator transport. At the same time, the entanglement entropy of stabilizer states can be computed exactly from their stabilizer structure~\cite{fattal2004}. Clifford circuits therefore provide a setting in which both entanglement growth and operator spreading can efficiently be computed and compared directly. 
Correspondingly, the OTOC reduces to a binary indicator of whether the evolved Pauli operator $W(t)$ has developed nontrivial Pauli support on site $x$, as determined by whether it commutes or anticommutes with $V_x$, a property we exploit in the diagnostics defined below.

\section{Model}\label{sec:Model}

We consider a one-dimensional quantum circuit acting on a chain of $N$ qubits, labeled by sites
$x\in[N]:=\{1,2,\ldots,N\}$. The elementary building block is a fixed $n$-qubit Clifford unitary that prepares a chosen graph state, whose internal structure is illustrated in Fig.~\ref{fig:block-structure}~(a). 
Time evolution proceeds in discrete layers, with each layer consisting of several nonoverlapping applications of this same $n$-qubit gate placed at random positions along the chain, as shown schematically in Fig.~\ref{fig:block-structure}(b).

More precisely, let $\mathcal{G}$ denote a fixed graph on $n$ vertices, and let $\mathcal{U}_{\mathcal{G}}\in \mathrm{SU}(2^n)$ be a Clifford unitary that prepares the corresponding graph state from the computational basis state,
\begin{equation}\label{eq:unitary-prepares-graphstate}
    |\mathcal{G}\rangle = \mathcal{U}_{\mathcal{G}} |0\rangle^{\otimes n},
\end{equation} 
where $|\mathcal{G}\rangle$ is the corresponding $n$-qubit graph state defined in Eq.~\eqref{eq:graph state},~\cite{hein2006graphstatesapplications,schlingemann2003clusterstatesalgorithmsgraphs}. The choice of $\mathcal{G}$ specifies the internal structure of the local circuit block and is held fixed throughout the time evolution.
This structure is illustrated in Fig.~\ref{fig:block-structure}.

At each discrete time step $t\in\mathbb{N}$, the global unitary $\mathscr{U}_t$ is given by a product of $r$ identical block unitaries acting on disjoint subsets of qubits,
\begin{equation}
    \mathscr{U}_t := \prod_{i=1}^{r} \mathcal{U}_{\mathcal{G}}\big[\mathcal{N}_{x_i}\big],
\end{equation}
where $\mathcal{U}_{\mathcal{G}}[\mathcal{N}_{x_i}]$ denotes the action of $\mathcal{U}_{\mathcal{G}}$ on the set of sites
\begin{equation}
    \mathcal{N}_{x_i} := \{x_i, x_i+1, \ldots, x_i+n-1\}.
\end{equation}

Throughout most of the paper we impose periodic boundary conditions, so that site indices are understood modulo $N$. 
We denote by $r_m=\lfloor N/n\rfloor$ the maximal number of nonoverlapping length-$n$ blocks that can be placed in a single layer.
The positions $\{x_i\}_{i=1}^r$ are chosen independently and uniformly at random at each time step, subject to the hard constraint that the corresponding supports $\mathcal{N}_{x_i}$ do not overlap. In practice, this is implemented by randomly sampling a set of $r$ disjoint intervals of length $n$ from the periodic set of $N$ sites. We parameterize the density of applied blocks by the \emph{sparsity}
\begin{equation}
    \alpha := \frac{r}{r_m},
\end{equation}
which quantifies the fraction of qubits acted on by graph-state blocks in each circuit layer, and thus controls the density of local updates. Varying $\alpha$ interpolates between sparse and dense circuit dynamics. We keep $\alpha$ fixed as $N$ is varied. Unless stated otherwise, we set $\alpha=1/2$ in the main text. 

The state of the system evolves according to 
\begin{equation}
     |\psi_t\rangle = \mathscr{U}_t |\psi_{t-1}\rangle\, ,
\end{equation}
starting from the product state $|\psi_0\rangle = |0\rangle^{\otimes N}$.
Since both the initial state and all circuit layers are Clifford, the state $|\psi_t\rangle$ remains a stabilizer state at all times~\cite{gottesmanthesis,Aaronson-Gottesman,hein-graphstates}. While $|\psi_t\rangle$ is not generally a graph state in the canonical $\prod \mathrm{CZ}\,|+\rangle^{\otimes N}$ form (Eq.~\eqref{eq:graph state}), it is always LC-equivalent to a graph state. In the following, we nevertheless use the language of \emph{injecting a graph-state block} to emphasize that the elementary local update is the fixed $n$-qubit graph-state preparation unitary $\mathcal{U}_{\mathcal{G}}$.

This construction defines an ensemble of random Clifford circuits that is fully specified by four ingredients: \emph{the block size $n$, the choice of graph $\mathcal{G}$ as the building block, the sparsity $\alpha$, and the system size $N$}. 
The circuit dynamics therefore depends on the choice of $\mathcal{G}$. 
Graph states within the same LC-equivalence class are locally unitary equivalent as quantum states. However, in our setting graph states enter through their preparation circuits, which are used repeatedly as dynamical building blocks. LC-equivalence of the target state does not in general imply identical circuit dynamics when these preparation unitaries are embedded into a layered architecture. We therefore fix one canonical preparation circuit per LC class (chosen for simplicity of implementation) and compare the resulting dynamics across these representatives.

\section{Diagnostics and velocity extraction}\label{sec:diagnostics}

In this section we define the diagnostics used to characterize entanglement growth and operator spreading in the circuit ensemble introduced in Sec.~\ref{sec:Model}, and specify how the corresponding velocities are extracted from numerical data. Throughout, all quantities are evaluated for fixed system size $N=200$ unless stated otherwise, and results are averaged over independent circuit realizations.

\subsection{Bipartite entanglement and entanglement velocity} 

We quantify entanglement growth using the bipartite von Neumann entropy~\cite{Amico-2008-many-body,RevModPhys.82.277}. For a pure state $|\psi_t\rangle$ evolving under the circuit, we consider a bipartition of the chain into a region $A=[1,\ldots,\ell]$ and its complement $\bar A$, and define
\begin{equation}
    S_A(t) = - \mathrm{Tr}\!\left[ \rho_A(t)\,\log \rho_A(t) \right],
\end{equation}
where $\rho_A(t)=\mathrm{Tr}_{\bar A}\big(|\psi_t\rangle\langle\psi_t|\big)$. Unless stated otherwise, we fix $\ell=N/2$ and use the natural logarithm throughout; changing the logarithm base rescales entropies by a constant factor and does not affect velocity comparison.

Because the circuit dynamics are Clifford and the initial state is a stabilizer state, the evolving state $|\psi_t\rangle$ remains a stabilizer state at all times~\cite{gottesmanthesis,Aaronson-Gottesman,hein-graphstates}. This allows the entanglement entropy to be computed efficiently using standard stabilizer techniques by counting the number of independent stabilizers fully supported within region $A$~\cite{fattal2004,hein-graphstates} (see also Appendix~\ref{appendix:stabilizer_entropy} for details). 

For the random circuit ensembles considered here, the entanglement entropy exhibits a linear growth regime at early and intermediate times,
\begin{equation}
    S_A(t) \simeq v_E\, t,
\end{equation}
before saturating to a volume-law value proportional to $\ell$, as observed in a wide range of random circuit models~\cite{Nahum-PhysRevX-2017,Zhou-Nahum-2019,Haferkamp_2022,vonKeyserlingk2018operator}. We extract the entanglement velocity $v_E$ by performing a linear fit to $S_A(t)$ over a time window chosen within the linear growth regime, after initial transients and before saturation.

\subsection{Out-of-time-ordered correlators and butterfly velocity}

To characterize operator spreading, we use out-of-time-ordered correlators defined in Eq.~\eqref{eq:OTOC}, constructed from local Pauli operators~\cite{Larkin1969,Swingle2018Unscrambling,Nahum-PhysRevX-2018}.
In Clifford circuits, this choice allows the OTOC to be evaluated exactly and interpreted directly in terms of operator support.
Specifically, we choose $W=X_{N/2}$, a Pauli-$X$ operator acting on the central site of the chain, and $V_x=Y_x$, a Pauli-$Y$ operator acting on site $x$.
We work in the Heisenberg picture, where operators evolve in time while the state remains fixed. Under the circuit dynamics generated by the unitary $\mathscr{U}_t$, the time-evolved operator is given by
\begin{equation}
    W(t) = \mathscr{U}_t^\dagger W \mathscr{U}_t\, ,
\end{equation}
which is the standard Heisenberg evolution for discrete-time quantum circuits.

For Pauli operators at infinite temperature, the general definition in Eq.~\eqref{eq:OTOC}
can be rewritten in an equivalent and computationally convenient form as

\begin{equation}
    C(x,t) = \frac{1}{2}\left(1 - \frac{1}{2^N}\mathrm{Tr}\!\left[ W(t)\,V_x\,W(t)\,V_x \right]\right)\, ,
    \label{eq:otoc_def}
\end{equation}
which follows from rewriting the squared commutator in Eq.~\eqref{eq:OTOC} in terms of operator products and using the normalization of the infinite-temperature trace (see Appendix~\ref{appendix:otoc} for details).

For Pauli operators, this quantity takes the value $C(x,t)=0$ if $W(t)$ and $V_x$ commute, and $C(x,t)=1$ if they anticommute. Because Clifford evolution maps Pauli operators to Pauli strings~\cite{gottesmanthesis,Aaronson-Gottesman,hein-graphstates}, $W(t)$ remains a Pauli string (spatial support can generally grow with time), and the OTOC defined in Eq.~\eqref{eq:otoc_def} provides a sharp indicator of whether the evolved operator has nontrivial Pauli support on site $x$ (see also Appendix~\ref{appendix:otoc} for details). 

For a fixed circuit realization, $C(x,t)$ therefore exhibits a sharp light-cone structure. To obtain smooth spatiotemporal profiles, we average $C(x,t)$ over a large number of
independent circuit realizations, yielding a continuous front that propagates outward from the center of the chain. The resulting averaged OTOC is used to extract the butterfly velocity $v_B$, defined as the speed at which the operator front advances.

Operationally, we determine $v_B$ by tracking the position $x(t)$ at which the averaged OTOC crosses a fixed threshold value and performing a linear fit to the resulting trajectory. The threshold value is chosen consistently across circuit realizations.

\subsection{Averaging and comparison across graph-state blocks}

All reported values of $v_E$ and $v_B$ are obtained by averaging over a sufficient number of independent circuit realizations for each choice of graph-state block, block size $n$, and sparsity $\alpha$ such that the quantities are well converged.
The number of realizations used in each case is specified in the corresponding figure captions.
When comparing different graph states, all global circuit parameters—including $N$, $\alpha$, boundary conditions, fitting windows, and averaging procedures—are held fixed.
Observed differences in entanglement and butterfly velocities can therefore be attributed directly to the internal structure of the graph-state blocks, rather than to changes in circuit architecture or randomness.
In the following section, we use these diagnostics to show that graph-state blocks that are inequivalent under local Clifford transformations generate distinct entanglement and operator-spreading dynamics.

\section{Results}\label{sec:results}

In this section we present numerical results characterizing entanglement growth and operator spreading in the 
random Clifford circuits built from graph-state blocks defined in Sec.~\ref{sec:Model}. Our analysis is organized around two complementary dynamical quantities: the entanglement velocity $v_E$, extracted from the growth of bipartite entanglement entropy, and the butterfly velocity $v_B$, extracted from the propagation of out-of-time-ordered correlators. By comparing circuits constructed from locally Clifford-inequivalent graph-state blocks while keeping all global circuit parameters fixed, we isolate how the internal structure of the blocks controls coarse-grained dynamical behavior. The corresponding codebase is available in Ref.~\cite{Github}.

We begin by analyzing the growth of bipartite entanglement under the random Clifford circuits introduced in Sec.~\ref{sec:Model}. Our primary goal is to determine how the internal structure of the graph-state blocks influences the entanglement velocity $v_E$, when all global features of the circuit architecture are held fixed.

Unless stated otherwise, we consider a system of size $N=200$ with periodic boundary conditions and sparsity $\alpha=1/2$. Entanglement is quantified by the von Neumann entropy $S_A(t)$ of a contiguous subsystem $A=[1,\ldots,N/2]$, as defined in Sec.~\ref{sec:diagnostics}, and results are averaged over a large number of independent circuit realizations.

\subsection{Entanglement growth and entanglement velocity}
\label{subsec:entanglement_growth}

\begin{figure*}[t]
\centering
\includegraphics[width=1.0\linewidth]{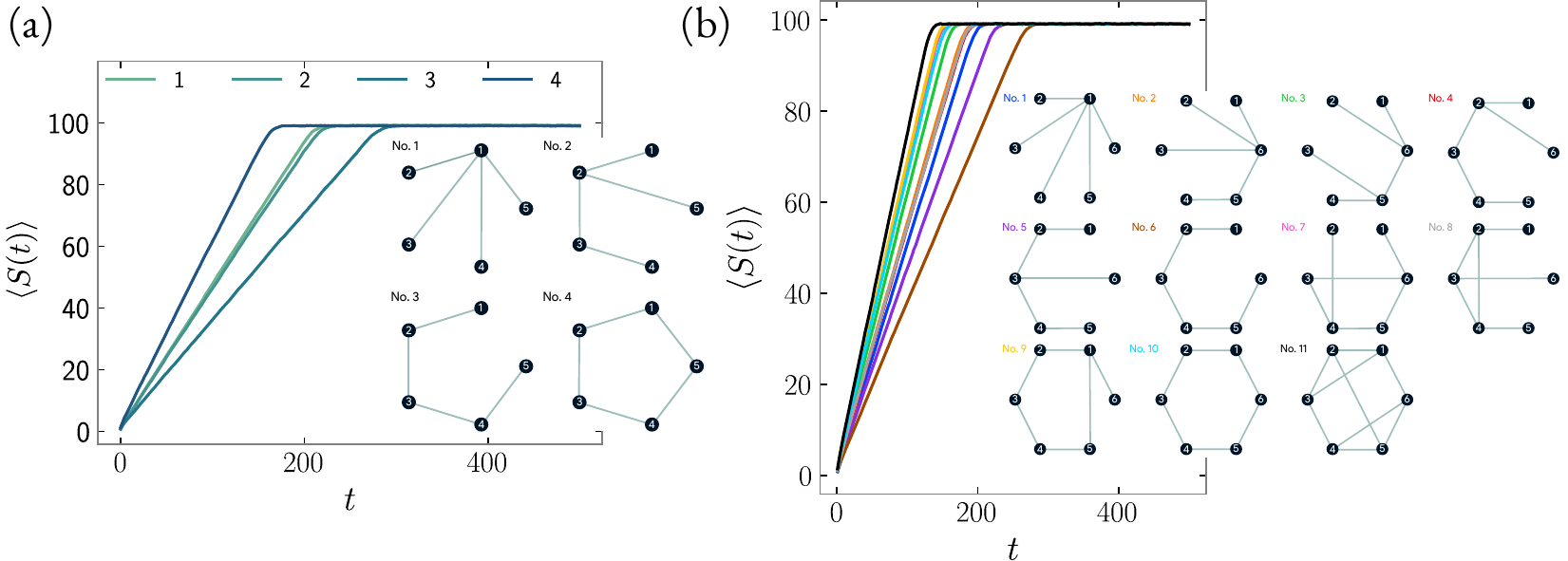}
\caption{
Bipartite entanglement growth for random Clifford circuits built from LC–inequivalent (a) five-qubit (b) six-qubit graph-state blocks.
The von Neumann entropy $S_A(t)$ of a half-chain subsystem $A=[1,\ldots,N/2]$ is shown as a function of circuit depth $t$ for $N=200$ and sparsity $\alpha=1/2$, averaged over independent circuit realizations.
Each curve corresponds to a distinct LC-inequivalence class of graph states of the corresponding block size.
All curves exhibit linear growth followed by saturation to a volume-law value, but with block-dependent slopes defining distinct entanglement velocities $v_E$. 
}
\label{fig:entanglement-g5&6}
\end{figure*}

Importantly, the \emph{rate} of entanglement growth depends strongly on the choice of graph-state block. Figure~\ref{fig:entanglement-g5&6}~(a) shows representative entanglement growth curves for circuits constructed from four LC-inequivalent five-qubit graph states (also see Figure~\ref{fig:5q-graph}).
While the qualitative behavior is the same in all cases, the slopes of the linear growth regimes—and hence the extracted entanglement velocities $v_E$—differ markedly between blocks. 
Representative entanglement growth curves for five- and six-qubit graph-state blocks are shown in Fig.~\ref{fig:entanglement-g5&6}.

The corresponding values of $v_E$ are also summarized in Table~\ref{tab:g456}. These results demonstrate that, even within a fixed Clifford circuit architecture with identical randomness in block placement, the internal entanglement structure of the local building block leaves a persistent imprint on coarse-grained entanglement dynamics.

We have verified that this behavior is robust under variation of the sparsity parameter $\alpha$: increasing $\alpha$ increases the rates of entanglement growth and operator spreading, while preserving the relative ordering across graph-state blocks. These results are presented in Appendix~\ref{appendix:sparsity}.

\begin{table}
\centering
\caption{\label{tab:g456} 
Entanglement and scrambling velocities for random Clifford circuits constructed from LC-inequivalent graph-state blocks of size $n=4,5,6$. Shown are the entanglement velocity $v_E$, butterfly velocity $v_B$, average height $\gamma$, and the block-level connectivity measure $\wp$ defined in the text.
All values are extracted from circuits with $N = 200$ and sparsity $\alpha = 1/2$ using identical fitting and averaging procedures, see~\cite{Github} for details. 
Rows corresponding to graph-state blocks realizing absolutely maximally entangled (AME) states are highlighted (for $n=5$: graph 4; for $n=6$: graph 11).
}
\vspace{0.2em}

\rowcolors{-1}{}{gray!10}

\begingroup
\setlength{\tabcolsep}{2.0pt} 
\renewcommand{\arraystretch}{1.05} 
\small 

\begin{tabular}{ccccccc}
\toprule
$n$ & Graph & Shape & $v_E$ & $v_B$ & $\gamma$ & $\wp$ \\
\midrule
~~4~~ & 1  & \includegraphics[height=5.5em,valign=c]{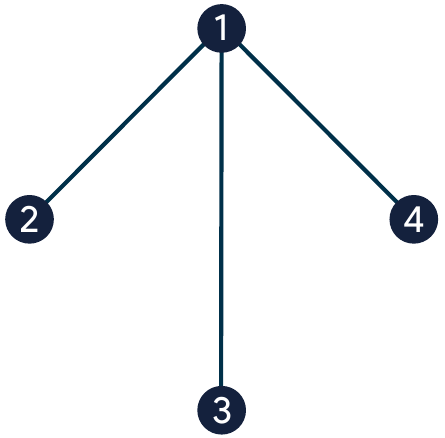} & ~~0.420~~ & ~~0.628~~ & ~~1~~ & ~~6~~ \\
4 & 2  & \includegraphics[height=5.5em,valign=c]{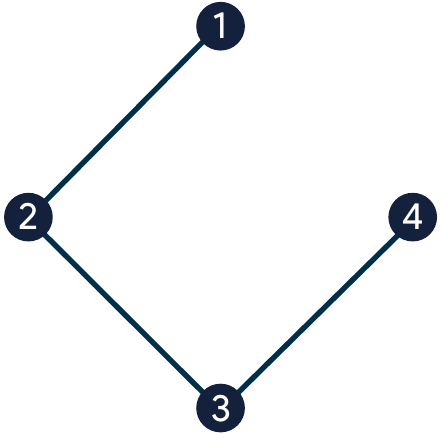} & 0.318 & 0.327 & 1 & 3 \\
\midrule
5 & 1  & \includegraphics[height=5.5em,valign=c]{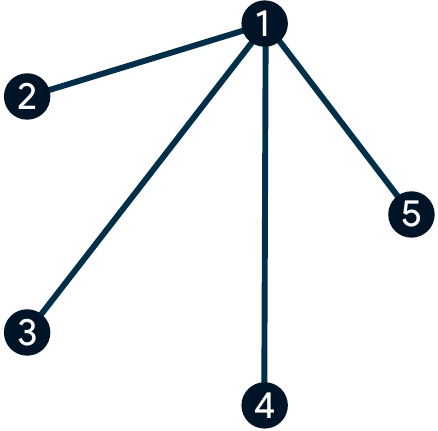} & ~~0.477~~ & ~~0.842~~ & ~~1.0~~ & ~~10~~ \\
5 & 2  & \includegraphics[height=5.5em,valign=c]{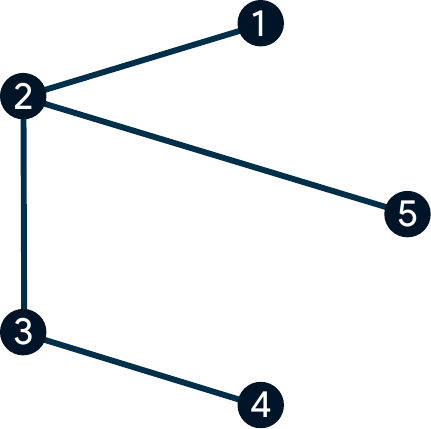} & 0.466 & 0.566 & 1.25 & 6 \\
5 & 3  & \includegraphics[height=5.5em,valign=c]{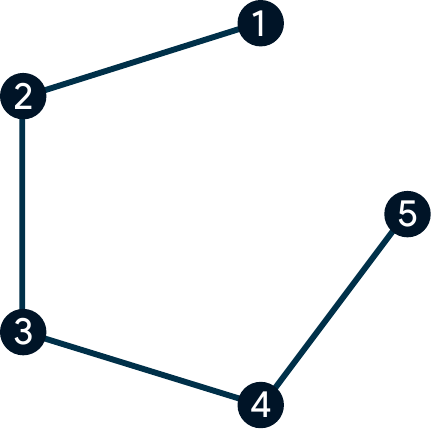} & 0.364 & 0.376 & 1.0 & 4 \\
\rowcolor{blue!20} 5 & 4  & \includegraphics[height=5.5em,valign=c]{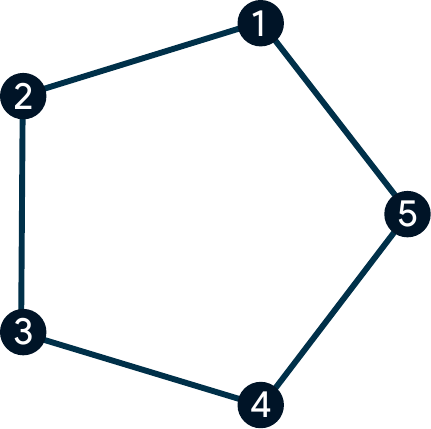} & 0.621 & 0.778 & 1.5 & 8 \\
\midrule
6 & 1  & \includegraphics[height=5.5em,valign=c]{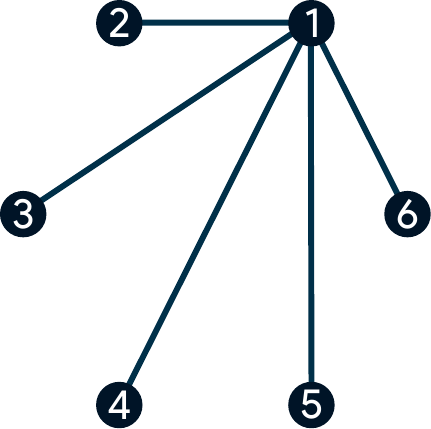} & 0.522 & 1.295 & 1.0 & 10 \\
6 & 2  & \includegraphics[height=5.5em,valign=c]{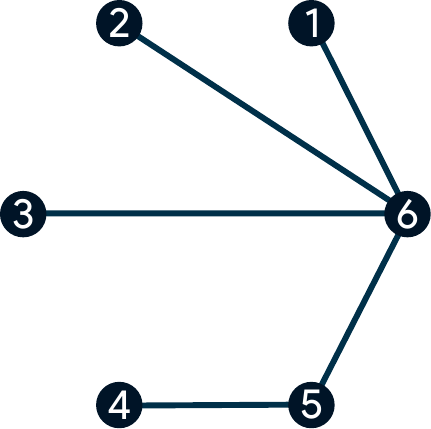} & 0.573 & 1.198 & 1.2 & 1 \\
6 & 3  & \includegraphics[height=5.5em,valign=c]{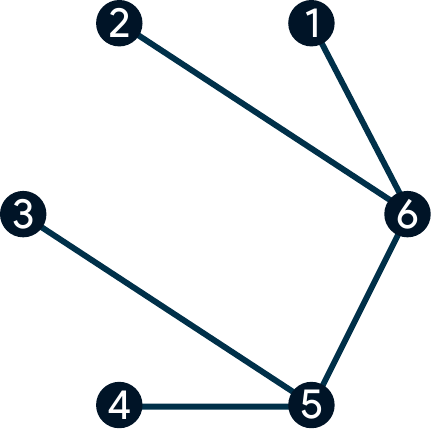} & 0.637 & 1.239 & 1.4 & 3 \\
6 & 4  & \includegraphics[height=5.5em,valign=c]{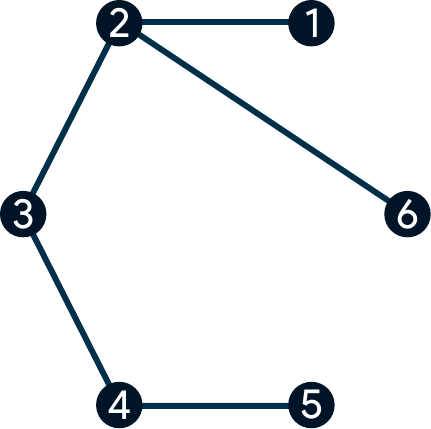} & 0.564 & 0.847 & 1.4 & 4 \\
\bottomrule
\end{tabular}
\endgroup
\end{table}

\begin{table}
\centering

\rowcolors{-1}{}{gray!10}

\begingroup
\setlength{\tabcolsep}{2.0pt} 
\renewcommand{\arraystretch}{1.05} 
\small

\begin{tabular}{ccccccc}
~~6~~ & ~~5~~  & ~~\includegraphics[height=5.5em,valign=c]{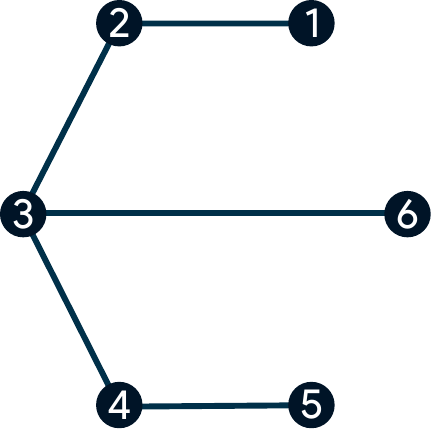}~~ & ~~0.469~~ & ~~0.654~~ & ~~1.2~~ & ~~4~~ \\
6 & 6  & \includegraphics[height=5.5em,valign=c]{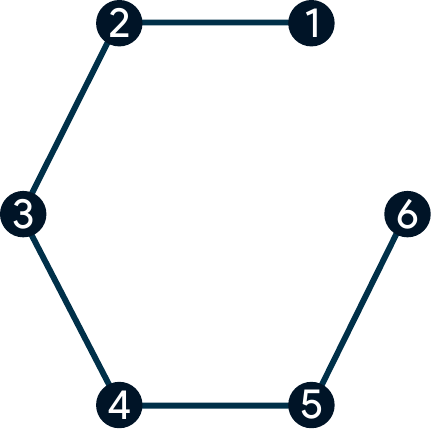} & 0.391 & 0.476 & 1.0 & 4 \\
6 & 7  & \includegraphics[height=5.5em,valign=c]{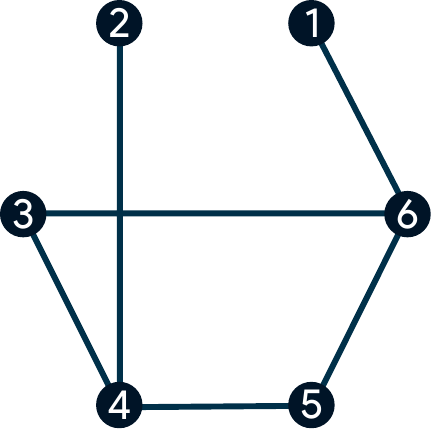} & 0.706 & 1.15  & 1.6 & 4 \\
6 & 8  & \includegraphics[height=5.5em,valign=c]{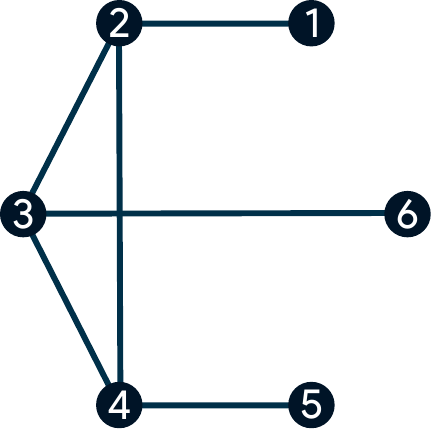} & 0.567 & 0.797 & 1.4 & 6 \\
6 & 9  & \includegraphics[height=5.5em,valign=c]{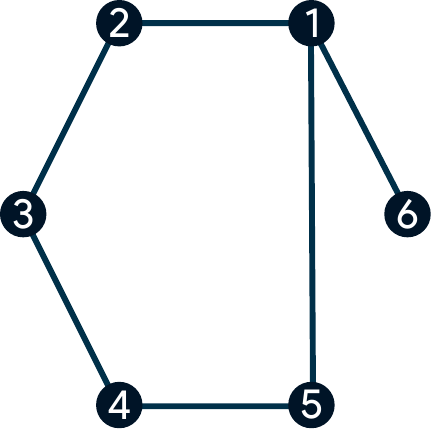} & 0.716 & 1.219 & 1.6 & 8 \\
6 & 10 & \includegraphics[height=5.5em,valign=c]{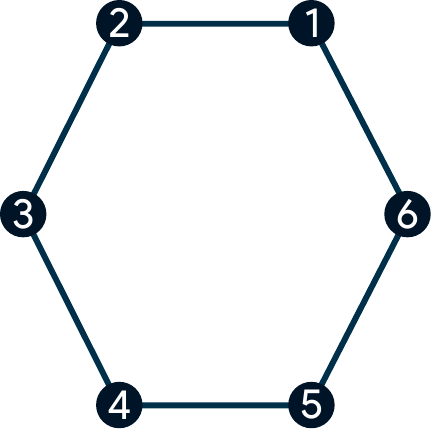} & 0.685 & 1.064 & 1.6 & 4 \\
\rowcolor{blue!20} 6 & 11 & \includegraphics[height=5.5em,valign=c]{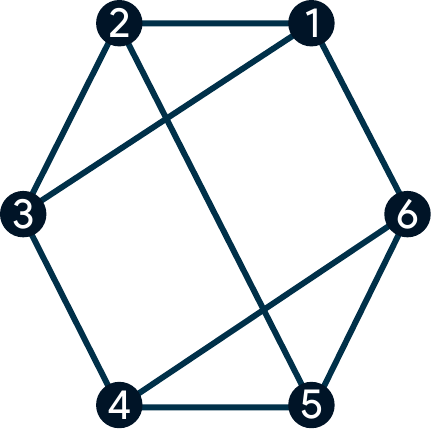} & 0.779 & 1.251 & 1.8 & 9 \\
\bottomrule
\end{tabular}
\endgroup
\end{table}

\paragraph{Linear entanglement growth and block-dependent velocities.}

For all graph-state blocks considered, the entanglement entropy exhibits a robust linear growth regime at early and intermediate times, followed by saturation to a volume-law value proportional to the subsystem size. This behavior is consistent with previous studies of random unitary and Clifford circuits~\cite{Nahum-PhysRevX-2017,Haferkamp_2022}, and shows that circuits built from graph-state blocks exhibit the characteristic linear entanglement growth associated with scrambling dynamics.

\paragraph{Block entangling capacity and the height function.}

To understand the observed hierarchy of entanglement velocities, we introduce a block-level measure that characterizes how entanglement is distributed across internal bipartitions of the graph-state block. For a given $n$-qubit graph state $|\mathcal{G}\rangle$, we define the \emph{height function} $h(x)$ as the bipartite entanglement entropy of $|\mathcal{G}\rangle$ across the cut separating sites $\{1,\ldots,x\}$ from $\{x+1,\ldots,n\}$, for $x=1,\ldots,n-1$ (see also~\cite{Li_2022}). Motivated by the fact that block placements are uniformly random and that a cut through the circuit can intersect a block at any internal position with equal probability, we consider the average height
\begin{equation}\label{eq:average-height}
    \gamma = \frac{1}{n-1}\sum_{x=1}^{n-1} h(x).
\end{equation}
The quantity $\gamma$ provides a coarse measure of the intrinsic entangling capacity of the block, averaged over all possible internal cuts.
As shown in Table~\ref{tab:g456}, blocks with larger $\gamma$ generally exhibit larger entanglement velocities $v_E$. 

\begin{figure}
    \centering
    \includegraphics[width=0.85\linewidth]{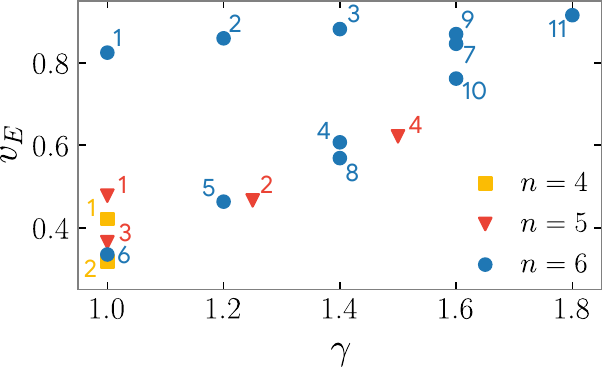}
    \caption{
Entanglement velocity $v_E$ as a function of the average height $\gamma$ for graph-state blocks of sizes $n=4,5,6$. Each point corresponds to a locally Clifford–inequivalent graph state (numbers indicate graph labels in Table~\ref{tab:g456}). For the block sizes where AME graph states occur here ($n=5,6$), the AME blocks attain the largest values of both $\gamma$ and $v_E$.
 The presence of two branches shows that $\gamma$ alone does not uniquely determine the entanglement velocity. Different markers denote different block sizes.
}
    \label{fig:ve-gamma}
\end{figure}

To visualize the relationship between the internal entanglement structure of the block and the entanglement velocity, Fig.~\ref{fig:ve-gamma} shows the extracted values of $v_E$ as a function of the average height $\gamma$ for all graph-state blocks with $n=4,5,6$. The data do not collapse onto a single curve. Instead, the points organize into two distinct branches, indicating that the average height $\gamma$ alone does not fully determine the entanglement velocity. Graph states with the same values of $\gamma$ can thus produce different $v_E$, reflecting differences in how entanglement is distributed across the internal bipartitions of the block. This observation suggests that additional structural features of the block also influence the dynamics. In the next subsection we show that operator spreading is governed by a complementary descriptor based on the internal connectivity of the block. The numerical values are summarized in Table~I, while the correlations are visualized in Fig.~\ref{fig:ve-gamma}.

In particular, the graph-state block realizing an absolutely maximally entangled (AME) state~\cite{AME-review,Goyeneche_2018}—corresponding to graph number~4 for $n=5$ in Fig.~\ref{fig:5q-graph} and Fig.~\ref{fig:entanglement-g5&6}~(a), graph number~11 for $n=6$ in Fig.~\ref{fig:entanglement-g5&6}(b)—attains both the largest average height and the fastest entanglement growth among the five-qubit blocks considered. 

\paragraph{Limits of a single-parameter description.}

While the average height $\gamma$ captures an important trend, it does not uniquely determine the entanglement velocity. Notably, there exist pairs of graph-state blocks with identical values of $\gamma$ but significantly different $v_E$. This indicates that entanglement growth in the full circuit is not governed solely by a single scalar measure of block entanglement, but also depends on how entanglement is distributed across different internal bipartitions.

This observation already suggests that entanglement growth and operator spreading may be controlled by distinct structural features of the block, motivating a complementary analysis based on operator dynamics. We return to this point in Sec.~\ref{subsec:otoc_results}, where we study operator spreading and butterfly velocities.

\paragraph{Dependence on block of graph states.}

The block-dependent hierarchy of entanglement and scrambling dynamics persists across block sizes. Repeating the analysis for $n=4,5$, and $6$, we find that different graph-state structures generate systematically different entanglement velocities $v_E$  and butterfly velocities $v_B$, as summarized in Table~\ref{tab:g456}. In particular, graph states with highly distributed internal entanglement—such as absolutely maximally entangled (AME) states when they exist—tend to produce the largest entanglement velocities $v_E$, while other structures with higher connectivity can yield larger butterfly velocities $v_B$. Additional results for $n=7$ are reported in Appendix~\ref{appendix:results for seven-qubit graph}. These observations show that the dynamical hierarchy observed here is not tied to a specific block size but reflects a robust feature of block-structured Clifford circuit dynamics.

\begin{figure*}[t]
    \centering
    \includegraphics[width=0.95\linewidth]{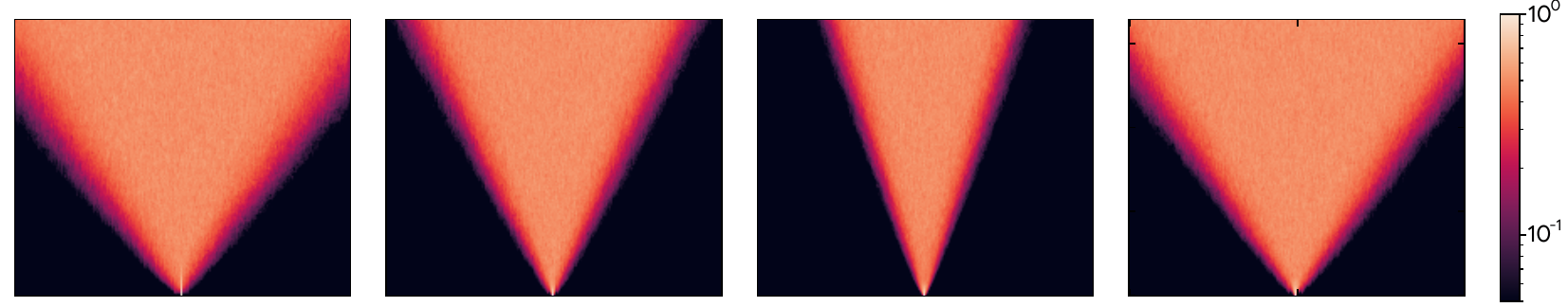}
    \caption{
Spatiotemporal profiles of the averaged out-of-time-ordered correlator $C(x,t)$ for random Clifford circuits constructed from LC–inequivalent five-qubit graph-state blocks.
The OTOC is computed from the commutator $[X_{N/2}(t),Y_x]$ for a system of size $N=200$ and sparsity $\alpha=1/2$, and averaged over independent circuit realizations.
In all cases, a ballistic light-cone structure is observed, with block-dependent slopes defining distinct butterfly velocities $v_B$.
Color scales are logarithmic to enhance contrast.
}
    \label{fig:otoc-g5}
\end{figure*}

\subsection{Operator spreading and butterfly velocity}
\label{subsec:otoc_results}

We now turn to operator dynamics and characterize the spreading of local operators under the same random Clifford circuits built from graph-state blocks. While entanglement growth probes how quantum correlations build up between subsystems, operator spreading captures how initially local information propagates through the system and becomes inaccessible to local measurements. These two diagnostics are closely related but not equivalent, and need not be governed by the same structural features of the circuit~\cite{Nahum-PhysRevX-2018,Xu-Swingle-PRXQ-2024,Zhou-Nahum-2019,Swingle2018Unscrambling}.

To probe operator spreading, we compute OTOCs as defined in Sec.~\ref{sec:diagnostics}. We focus on the Heisenberg evolution of a Pauli-$X$ operator initially localized at the center of the chain, $W=X_{N/2}$, and evaluate its commutation with local Pauli-$Y$ operators $V_x=Y_x$ at site $x$. Because the circuit dynamics are Clifford, the evolved operator $W(t)$ remains a Pauli string (i.e., a Pauli operator whose spatial support grows over time),
and provides a sharp indicator of whether $W(t)$ anticommutes with $Y_x$, i.e., whether the Pauli operator carried by $W(t)$ at site $x$ is $X$ or $Z$ (rather than $\mathbb{1}$ or $Y$). 

\paragraph{Light-cone structure and block-dependent butterfly velocities.}

Figure~\ref{fig:otoc-g5} shows the averaged OTOC, $C(x,t)$, for circuits constructed from four LC-inequivalent five-qubit graph-state blocks. In all cases, the OTOC exhibits a clear light-cone structure~\cite{Nahum-PhysRevX-2018,Xu-Swingle-PRXQ-2024}: outside a linearly expanding region, $C(x,t)$ remains close to zero, while inside the light cone it rapidly approaches its maximal value. The slope of this front defines the butterfly velocity $v_B$.

As in the case of entanglement growth, we find that the butterfly velocity depends strongly on the choice of graph-state block. While all blocks generate ballistic operator spreading, the extracted values of $v_B$ differ substantially between blocks, as summarized in Table~\ref{tab:g456}. These differences persist even though the global circuit architecture, randomness in block placement, and averaging procedures are identical.

\paragraph{Graph connectivity and operator transport.}

To understand the observed hierarchy of butterfly velocities, it is useful to recall how Pauli operators for qubit systems propagate under Clifford dynamics. Conjugation by a controlled-$Z$ gate maps local Pauli operators to products of Pauli operators on neighboring sites (e.g., $X_i \mapsto X_i Z_j$), while Hadamard gates exchange $X$ and $Z$. As a result, the growth of operator support is constrained by the pattern of entangling edges within the graph-state block.

\begin{figure}
    \centering
    \includegraphics[width=0.85\linewidth]{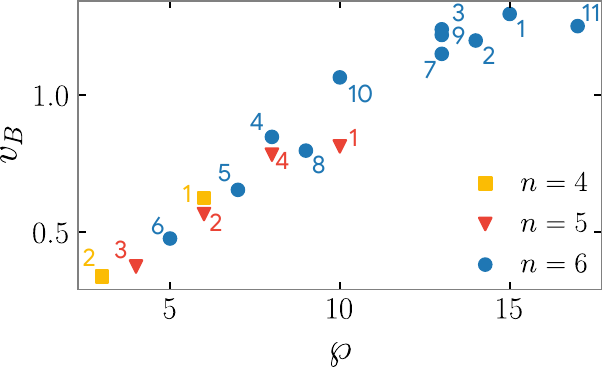}
    \caption{Butterfly velocity $v_B$ as a function of the connectivity measure $\wp$ for graph-state blocks of sizes $n=4,5,6$. As in Fig.~\ref{fig:ve-gamma}, each point corresponds to a locally Clifford–inequivalent graph state used as the circuit building block. A positive correlation is observed: blocks with larger $\wp$ generally exhibit faster operator spreading. As stated in the main text, the remaining scatter indicates that $\wp$ captures an important but not complete aspect of the operator transport dynamics. Different markers indicate different block sizes.
}
\label{fig:vb-conn}
\end{figure}

Motivated by this observation, we introduce a simple graph-theoretic measure that quantifies how efficiently a block can transport operator support across internal bipartitions. For a given graph $\mathcal{G}$, let $e_a$ denote the number of edges crossing the bipartition separating sites $\{1,\ldots,a\}$ from $\{a+1,\ldots,n\}$. We define the ordered graph connectivity
\begin{equation}\label{eq:connectivity}
    \wp = \sum_{a=1}^{n-1} e_a .
\end{equation}
This quantity counts the total number of entangling connections that allow operator support to propagate across different internal cuts of the block. Larger values of $\wp$ indicate that operator support can spread more efficiently across the block, facilitating faster spatial growth.
Fig.~\ref{fig:vb-conn} illustrates the descriptor $\wp$ for graph blocks $n=4,5$ and $6$.

As shown in Table~\ref{tab:g456}, blocks with larger values of $\wp$ generally exhibit larger butterfly velocities $v_B$. In particular, graph-state blocks that are highly connected across many bipartitions allow operator support to propagate more rapidly through the circuit. At the same time, $\wp$ does not uniquely determine $v_B$, reflecting the fact that successive Clifford conjugations can locally reconcentrate operator support and partially undo earlier spreading. Thus, as in the case of entanglement growth, operator spreading cannot be fully captured by a single scalar block parameter.

\paragraph{Complementarity with entanglement growth.}

Comparing the results for $v_E$ and $v_B$ reveals a clear separation of roles between the two block-level descriptors introduced in this paper $\gamma$ and $\wp$. The average height $\gamma$, Eq.~\eqref{eq:average-height}, captures how efficiently a block can generate entanglement across cuts, while the connectivity measure $\wp$, Eq.~\eqref{eq:connectivity}, captures how efficiently operator support can be transported.

This distinction is directly reflected in the dynamical data. In particular, one can identify pairs of graph-state blocks for which the ordering induced by $v_E$ differs from that induced by $v_B$. For example, for two six-qubit graph states number 1 and number 2 (see Table~\ref{tab:g456}), we find that $v_E\, (\text{no. 1})= 0.522 < v_E \, (\text{no. 2})= 0.573$, while at the same time $v_B\, (\text{no. 1})= 1.295 > v_B \, (\text{no. 2})= 1.198$. This inversion demonstrates that entanglement generation and operator spreading are governed by distinct structural features of the graph-state blocks. More generally, blocks with larger $\gamma$ tend to exhibit larger entanglement velocities $v_E$, while blocks with larger $\wp$ tend to exhibit larger butterfly velocities $v_B$. However, these two structural characteristics are not equivalent, and optimizing one does not necessarily optimize the other.

These findings demonstrate that entanglement growth and operator spreading, while correlated, are governed by distinct structural features of multipartite entangled circuit primitives. This distinction underlies the nonuniversal dynamical behavior observed in random Clifford circuits built from graph-state blocks.

Taken together, these results show that multipartite graph-state blocks define distinct dynamical classes in random Clifford circuits built from graph-state blocks, distinguished by how they generate entanglement and transport operators.

In the main text we present results for block sizes $n=4,5,$ and $6$, and report additional data for $n=7$ in the Appendix~\ref{appendix:results for seven-qubit graph}. Unless stated otherwise, all results are obtained for periodic boundary conditions; we verify that the qualitative hierarchy of dynamical behavior is robust to open boundary conditions, which are discussed separately in the Appendix~\ref{appendix:obc}. 

\subsection{Comparison across block sizes}

The preceding subsections show that both the entanglement velocity $v_E$ and the butterfly velocity $v_B$ depend strongly on the internal structure of the graph-state block. It is therefore natural to ask whether increasing the block size $n$ systematically enhances the rate of entanglement generation or operator spreading. Our numerical results indicate that this is not generally the case. Although larger blocks can potentially generate more entanglement within a single circuit layer, the observed velocities do not increase monotonically with $n$. For example, Table~\ref{tab:g456} shows that the five-qubit graph labeled 4 exhibits an entanglement velocity $v_E = 0.621$, while several six-qubit blocks yield smaller values of $v_E$, such as graph 6 with $v_E = 0.391$. A similar behavior is observed for the butterfly velocity $v_B$. In particular, the five-qubit graph labeled 2 has $v_B = 0.566$, whereas the six-qubit graph labeled 6 has $v_B = 0.476$.

These comparisons demonstrate that the coarse feature of block size alone does not determine the dynamical rates. Instead, the detailed internal structure of the graph-state block—captured by the descriptors $\gamma$ and $\wp$ introduced above—plays the dominant role in controlling entanglement growth and operator spreading in the circuit dynamics.

\section{Discussion and outlook}\label{sec:discussion}

In this paper we studied entanglement growth and operator spreading in a family of random Clifford circuits built from graph-state blocks.
In our model, the local circuit primitive is a fixed multipartite graph-state preparation unitary applied at random positions along a one-dimensional chain.
By comparing circuits constructed from LC–inequivalent graph-state blocks of sizes $n=4,5,6$ (with additional results for $n=7$; see Appendix~\ref{appendix:results for seven-qubit graph}), while keeping the global architecture and randomness fixed, we showed that coarse-grained dynamical properties are not universal within this class of models.
Instead, the internal structure of the circuit primitives induces a clear hierarchy of entanglement and scrambling rates. We also find that this hierarchy remains unchanged when the sparsity parameter $\alpha$ is varied, while the overall rates of entanglement growth and operator spreading increase with $\alpha$. Here, $\alpha$ controls the density of graph-state blocks applied in each circuit layer.

A central outcome of our analysis is the separation of roles played by two block-level descriptors.
The entanglement velocity $v_E$ is primarily governed by how entanglement is distributed across internal bipartitions of the block, as quantified by the average height $\gamma$.
In contrast, the butterfly velocity $v_B$ is controlled by how efficiently operator support can be transported across the block, as captured by the graph-connectivity measure $\wp$.
While these two quantities are correlated, neither alone fully determines the dynamics, highlighting the need for complementary structural descriptors to characterize multipartite circuit primitives.

Within this framework, graph-state blocks realizing absolutely maximally entangled (AME) states—or their closest stabilizer analogs when exact AME states do not exist—play a distinguished role because they maximize the internal distribution of entanglement across bipartitions. Consequently, when AME states exist they yield the largest entanglement velocities $v_E$  among the graph-state blocks studied here. By contrast, the largest butterfly velocities $v_B$ can arise from graph structures with stronger connectivity across bipartitions, such as GHZ-type graphs. 
More generally, graph-state blocks that generate entanglement most efficiently (large $v_E$) are not necessarily those that spread operators most rapidly (large $v_B$). This separation reflects the distinct structural requirements for entanglement generation and operator transport.
These observations highlight the complementary roles of the two descriptors $\gamma$ and $\wp$ in governing entanglement generation and operator transport.

Our results complement and refine the standard picture of randomness and scrambling in local quantum circuits.
Much of the existing literature focuses on circuits built from generic two-qubit gates drawn from Haar-random or design-forming ensembles, where coarse-grained quantities such as entanglement growth and operator spreading are often treated as universal features determined primarily by geometry and locality.
The present paper shows that when local dynamics are built from structured multipartite primitives, this universality can break down at the level of dynamical rates, even though linear entanglement growth and ballistic operator spreading persist.
In this sense, our findings align with recent studies demonstrating that strong pseudorandom behavior and complexity growth can emerge from non-Haar or constrained local dynamics, while emphasizing that the detailed structure of the local primitives still matters.

It is worth emphasizing that the restriction to Clifford circuits is a deliberate choice rather than a limitation.
Clifford dynamics allow exact tracking of stabilizers and Pauli operators, enabling a clean separation between entanglement generation and operator transport without finite-sampling noise or uncontrolled approximations.
This makes it possible to isolate how multipartite entanglement structure alone shapes coarse-grained dynamics.
While the quantitative values of $v_E$ and $v_B$ will generally differ in non-Clifford circuits, the qualitative distinction between entanglement capacity and transport capacity is expected to persist beyond the Clifford regime.

Several natural extensions of this work suggest themselves.
One direction is to interpolate between Clifford and non-Clifford dynamics by introducing a controlled density of non-Clifford gates, thereby probing the robustness of the observed hierarchy under increasing circuit complexity.
Another is to consider higher-dimensional geometries or alternative patterns of block placement, where connectivity and entanglement structure may interplay in new ways.
Finally, it would be interesting to connect the present dynamical diagnostics more directly to notions of circuit complexity growth and approximate unitary designs, further clarifying how structured local primitives shape the approach to effective randomness.

More broadly, our results illustrate that multipartite entangled states are not interchangeable as dynamical resources.
Even when embedded into otherwise identical random circuits, their internal structure can leave lasting signatures on how quantum information spreads and scrambles.
Understanding and exploiting this structure may prove useful in the design of quantum circuits, simulators, and architectures where entanglement and information flow are key resources.

\begin{acknowledgments}
We thank Wojciech Bruzda,
Jens Eisert, Otfried G\"uhne, Yien Liang, and Karol \.Zyczkowski for useful discussions.

The work of MF as well as, up until 31.08.2025, AB was supported by the Polish National Science Centre (NCN) grant 2021/42/E/ST2/00234. AB acknowledges support from United Kingdom Research and Innovation (UKRI) under the UK government’s Horizon Europe guarantee (EP/Y00468X/1) (since 01.09.2025).
HS acknowledges supports from Perimeter Institute for Theoretical Physics, a research institute supported in part by the Government of Canada through the Department of Innovation, Science and Economic Development Canada and by the Province of Ontario through the Ministry of Colleges and Universities.
ZR acknowledges support from the German Federal Ministry of Research, Technology and Space (BMFTR) through the Quantum Futur project "Quantum Secure Communication with Adaptive Layers and Ranks (Q-SCALAR)", Grant No. 13N17638, and support from the Equal Opportunity Program, Grant Line 2: Support for Female Junior Professors and Postdocs through Academic Staff Positions, 15th funding round, at Paderborn University.

\end{acknowledgments}

\vspace{0.5em}
\noindent \textit{Data availability.---} The data that supports the findings of this article are openly available at \cite{UJ/EJAAFN_2026}

\appendix 

\begin{table}
    \centering
    \caption{\label{tab:g7} Entanglement and scrambling velocities for random Clifford circuits constructed from LC-inequivalent seven-qubit graph-state blocks. In numerical simulation, we took system-size $N= 500$ and sparsity $\alpha = 1/2$.}
    \rowcolors{-1}{}{gray!10}
    \begin{tabular}{*6c}
        \toprule
        Graph & Shape & $v_E$ & $v_B$ & $\gamma$ & $\wp$ \\  
        \midrule
      1 & \includegraphics[height=5.5em, valign=c]{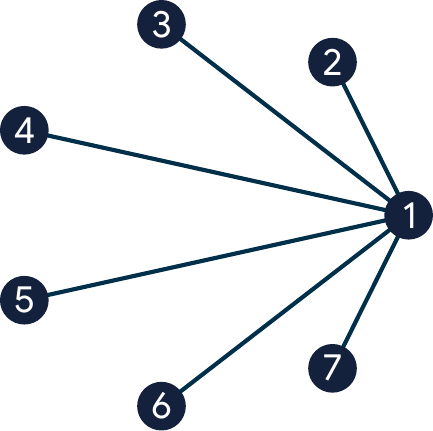} & ~~0.554~~ & ~~1.255 ~~ & ~~1.0~~ & ~~ 21 ~~  \\
      2 & \includegraphics[height=5.5em, valign=c]{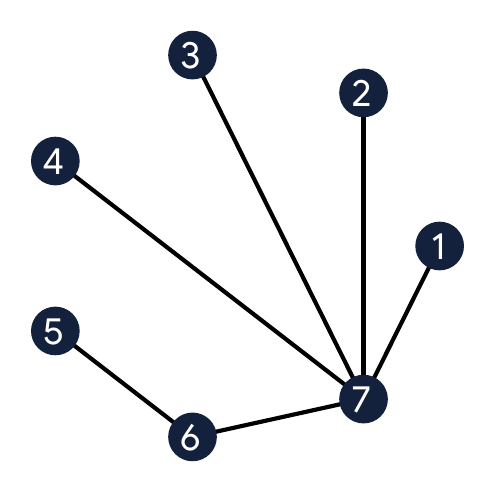} & 0.603 & 1.224  & 1.17 &  20\\
      3 & \includegraphics[height=5.5em, valign=c]{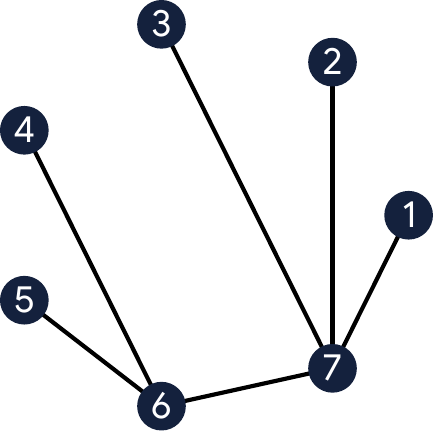} & 0.657 & 1.221 & 1.33 &  19\\
      4 & \includegraphics[height=5.5em, valign=c]{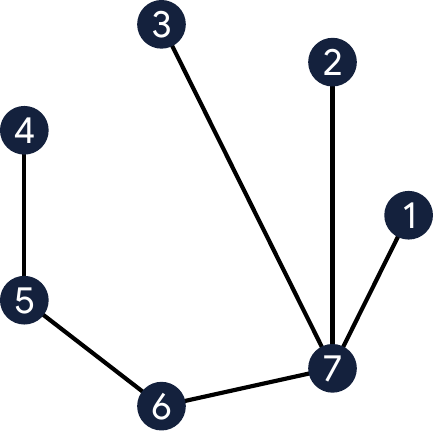} & 0.649 & 1.209& 1.33 &  18 \\
      5 & \includegraphics[height=5.5em, valign=c]{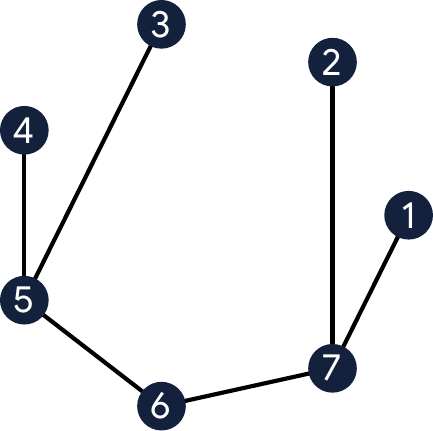} & 0.718 & 1.242& 1.5 &   16\\
      6 & \includegraphics[height=5.5em, valign=c]{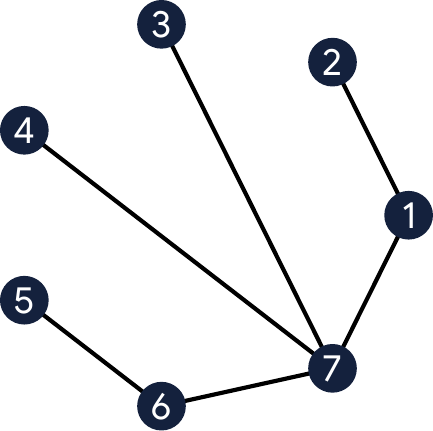} & 0.589 & 1.015& 1.17 &   16\\
      7 & \includegraphics[height=5.5em, valign=c]{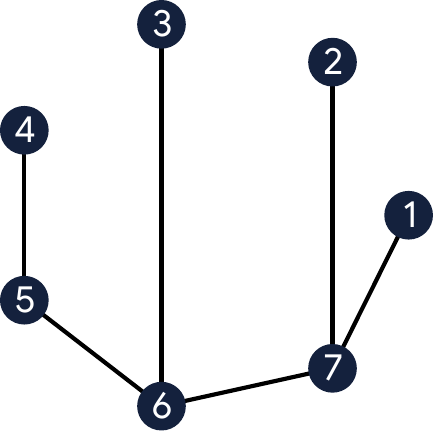} & 0.780 & 1.246& 1.67 &   17\\
      8 & \includegraphics[height=5.5em, valign=c]{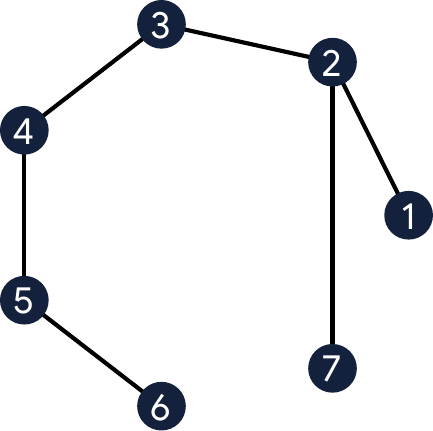} & 0.646 & 0.84 & 1.5 &   10\\
      9 & \includegraphics[height=5.5em, valign=c]{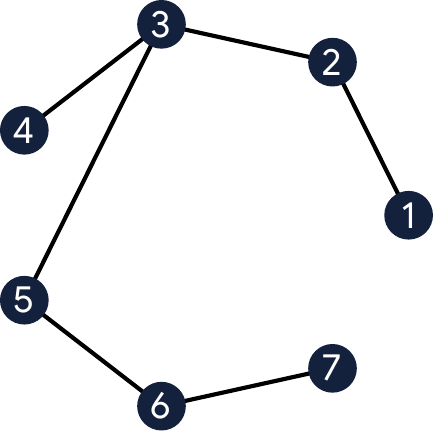} & 0.403 & 0.448 & 1.0 &   7\\
      10 & \includegraphics[height=5.5em, valign=c]{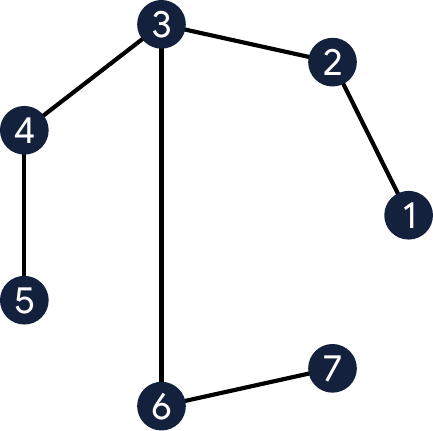} & 0.478 & 0.537 & 1.17 &  8 \\  
        \bottomrule
    \end{tabular}
\end{table}

\begin{table}
    \rowcolors{-1}{}{gray!10}
    \begin{tabular}{*6c}
        \toprule
    ~~11~~ & ~~\includegraphics[height=5.5em, valign=c]{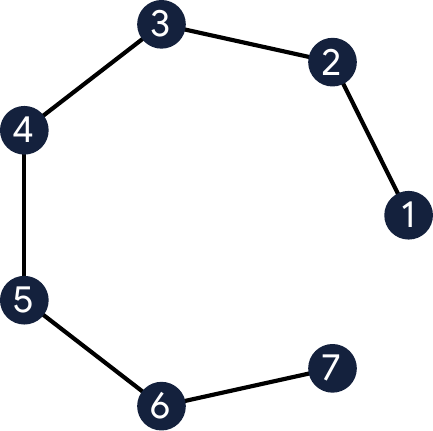}~~ & ~~0.426~~ & ~~0.474~~& ~~1.0~~ &   ~~6~~\\ 
      ~~12~~ & ~~\includegraphics[height=5.5em, valign=c]{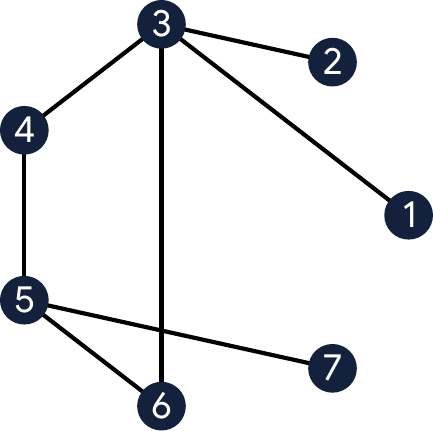}~~ & ~~0.380~~ & ~~0.501~~ & ~~1.0~~ &   ~~9~~ \\   
      13 & \includegraphics[height=5.5em, valign=c]{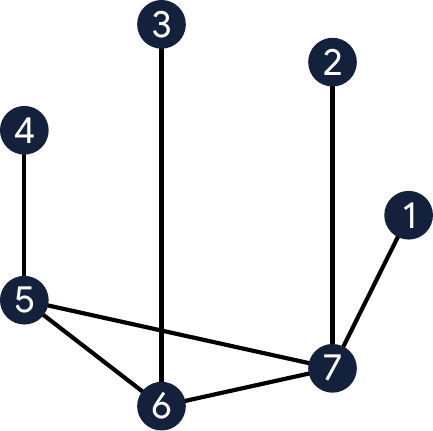} & 0.780 & 1.214 & 1.67 &  19 \\   
      14 & \includegraphics[height=5.5em, valign=c]{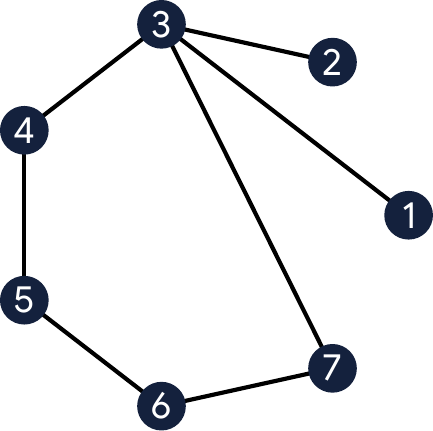} & 0.579 & 0.724 & 1.33 &  11 \\   
    15 & \includegraphics[height=5.5em, valign=c]{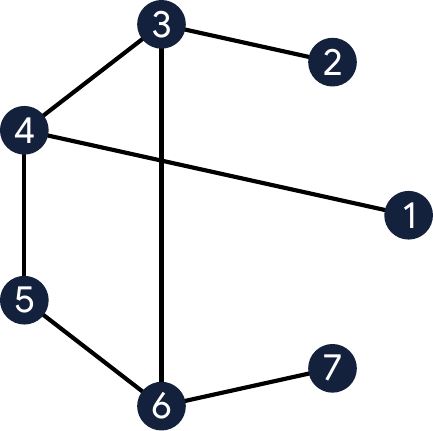} & 0.621 & 0.703 & 1.5 &   11\\  
      16 & \includegraphics[height=5.5em, valign=c]{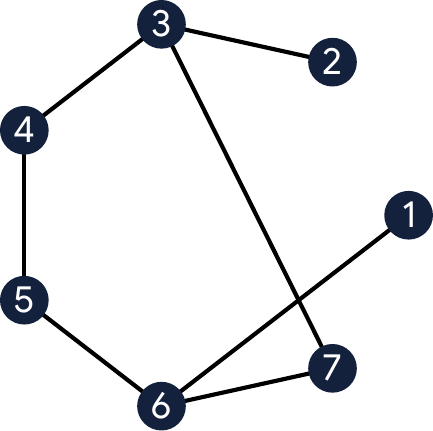} & 0.829 & 1.072 & 1.83 &  14 \\   
      17 & \includegraphics[height=5.5em, valign=c]{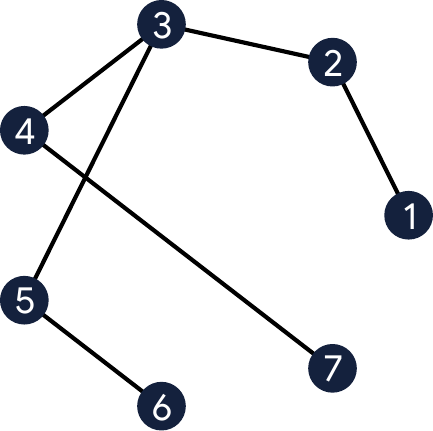} & 0.566 & 0.638& 1.33 &  9 \\   
      18 & \includegraphics[height=5.5em, valign=c]{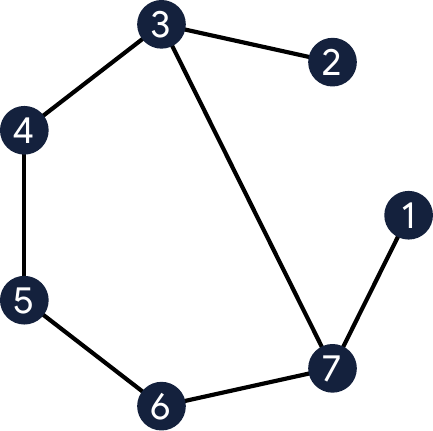} & 0.773 & 1.098 & 1.67 &  15 \\   
      19 & \includegraphics[height=5.5em, valign=c]{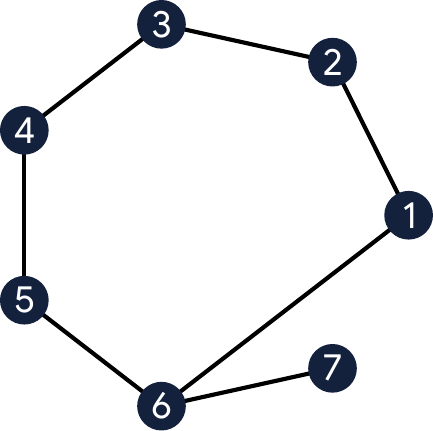} & 0.653 & 0.813 & 1.5 &   11\\   
    20 & \includegraphics[height=5.5em, valign=c]{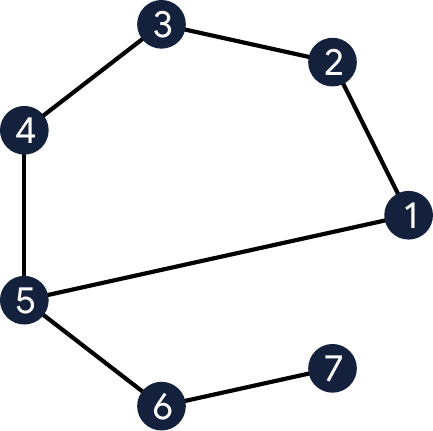} & 0.572 & 0.668 & 1.33 &  10 \\  
      21 & \includegraphics[height=5.5em, valign=c]{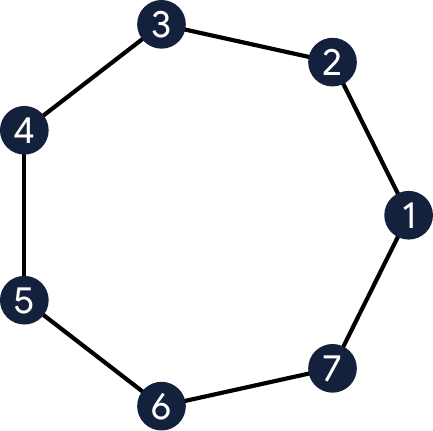} & 0.772 & 1.016 & 1.67 &  12 \\    
        \bottomrule
    \end{tabular}
\end{table}

\vspace{2cm}

\begin{table}
    \rowcolors{-1}{}{gray!10}
    \begin{tabular}{*6c}
        \toprule
    22 & \includegraphics[height=5.5em, valign=c]{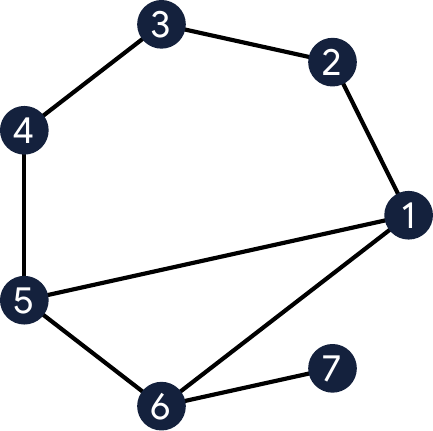} & 0.686 & 0.917 & 1.5 &  15 \\  
      23 & \includegraphics[height=5.5em, valign=c]{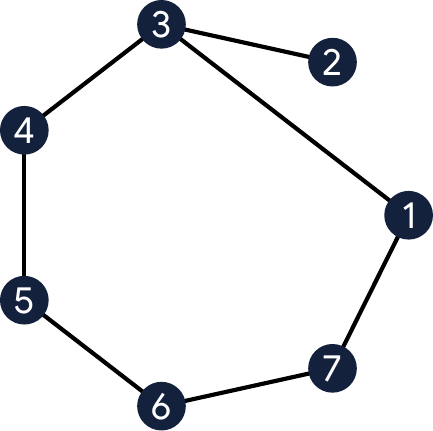} & 0.774 & 1.086 & 1.67 & 13  \\  
      ~~24~~ & ~~\includegraphics[height=5.5em, valign=c]{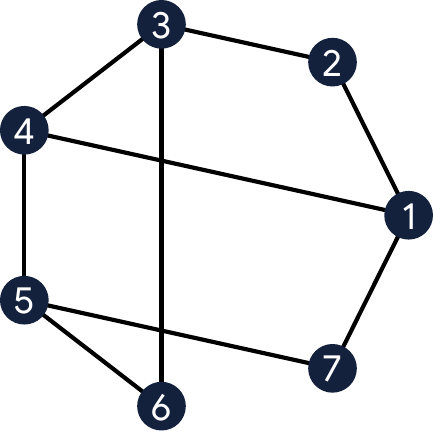}~~ & ~~0.833~~ & ~~1.19~~ & ~~1.83~~ &  ~~19~~ \\   
      25 & \includegraphics[height=5.5em, valign=c]{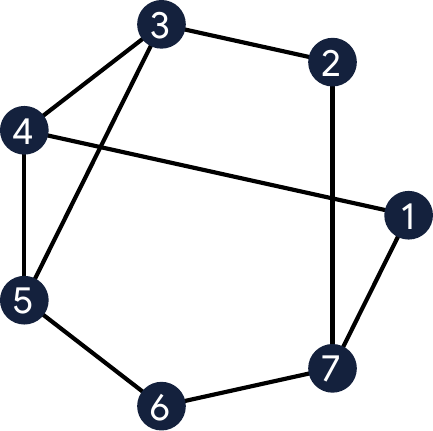} & 0.847 & 1.257 & 1.83 & 21  \\   
      26 & \includegraphics[height=5.5em, valign=c]{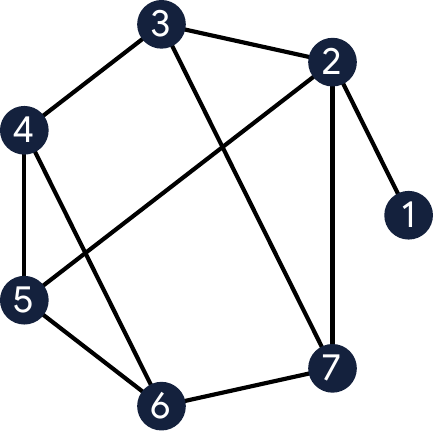} & 0.731 & 0.996 & 1.67 &  20 \\   
        \bottomrule
    \end{tabular}
\end{table}

\section{Results for seven-qubit graph}\label{appendix:results for seven-qubit graph}
In Table\,\ref{tab:g7}, we present numerically calculated quantities $(v_B, v_E, \gamma, \wp)$ for the LC-inequivalent seven-qubit graph states.

\section{Entanglement in Clifford circuits}\label{appendix:stabilizer_entropy}

An $n$ qubit stabilizer state, $|\psi \rangle$, is defined to be the unique $+1$ eigenvalue of $n$ independent Pauli operators $g_1,\ldots ,g_n$,
\begin{equation}
    |\psi\rangle\langle \psi | \,=\, \frac{1}{2^n} \prod_{i=1}^n (g_i + \mathbb{1}) \, =\, \frac{1}{2^n}\sum_{S\in \mathcal{S}} S
\end{equation}
where $\mathcal{S}$ is the stabilizer group generated by $g_1,\ldots g_n$ containing all Pauli operators $S$ which stabilize $|\psi\rangle $, i.e., $S|\psi\rangle = |\psi\rangle$.

The entanglement entropy of a stabilizer state in a region $A$ can be calculated by counting the number of stabilizers in the stabilizer group that are completely contained in $A$, i.e., that act trivially on its complementary. Then, the entanglement entropy is given by 
\begin{equation}
    S_A = n_A  -  \log_2 |\mathcal{S}_A|
\end{equation}
where $n_A$ is the number of qubits contained in $A$ and $|\mathcal{S}_A|$ is the size of the subgroup of stabilizers contained in $A$. This stabilizer formalism allows efficient and exact evaluation of entanglement entropy during the Clifford circuit evolution used in this work.

\section{Out-of-time-ordered correlators in Clifford circuits}
\label{appendix:otoc}

In this appendix, we provide additional details on the definition, normalization, and evaluation of the out-of-time-ordered correlator (OTOC) used in the main text. OTOCs and related squared-commutator diagnostics are widely used to characterize operator spreading and scrambling in quantum many-body systems; for general reviews and conventions, see Refs.~\cite{Xu-Swingle-PRXQ-2024,Nahum-PhysRevX-2018}, while the original idea of using out-of-time-order operator growth diagnostics goes back to Larkin and Ovchinnikov~\cite{Larkin1969}.

\subsection{Definition, normalization, and simplification for Pauli operators}

For two operators $W$ and $V_x$, the infinite-temperature OTOC is commonly written as
\begin{equation}
    F(x,t) = \frac{1}{2^N}\mathrm{Tr}\!\left[ W(t)\,V_x\,W(t)\,V_x \right],
\end{equation}
where
\begin{equation}
    W(t)=\mathscr{U}_t^\dagger W \mathscr{U}_t
\end{equation}
is the Heisenberg-evolved operator. In the literature, one also frequently considers the squared commutator
\begin{equation}
    C_{\mathrm{comm}}(x,t)=\frac{1}{2\cdot 2^N}\,
    \mathrm{Tr}\!\left([W(t),V_x]^\dagger [W(t),V_x]\right),
\end{equation}
because it is manifestly nonnegative and directly measures the extent to which the two operators fail to commute, making it a natural diagnostic of operator spreading; see, e.g., Refs.~\cite{Xu-Swingle-PRXQ-2024,Larkin1969}.

For Hermitian unitary operators such as Pauli operators, one has $W(t)^2=V_x^2=\mathbb{1}$, and the squared commutator can be written in terms of the OTOC. Expanding,
\begin{align}\nonumber
    &[W(t),V_x]^\dagger [W(t),V_x]\\
    &= \left( V_xW(t)-W(t)V_x \right) \left(W(t)V_x-V_xW(t)\right) \nonumber\\
    &= 2\mathbb{1}-W(t)V_xW(t)V_x - V_xW(t)V_xW(t)\ .\nonumber
\end{align}
Using cyclicity of the trace,
\begin{equation}
    \mathrm{Tr}\!\left[V_xW(t)V_xW(t)\right]
    =
    \mathrm{Tr}\!\left[W(t)V_xW(t)V_x\right],
\end{equation}
and therefore
\begin{equation}
    C_{\mathrm{comm}}(x,t)
    =
    1-\frac{1}{2^N}\mathrm{Tr}\!\left[W(t)V_xW(t)V_x\right].
\end{equation}
In the main text we use the rescaled quantity
\begin{equation}
    C(x,t)=\frac{1}{2}\left(1-\frac{1}{2^N}\mathrm{Tr}\!\left[W(t)V_xW(t)V_x\right]\right),
\end{equation}
which is Eq.~\eqref{eq:otoc_def} of the main text. This normalization is convenient in the present Clifford setting because it takes the binary values $0$ and $1$ for commuting and anticommuting Pauli operators, respectively.

Indeed, when $W(t)$ and $V_x$ are Pauli operators, they either commute or anticommute:
\begin{align}
    W(t)V_x &= +V_xW(t), \\
    W(t)V_x &= -V_xW(t).
\end{align}
In these two cases one obtains
\begin{equation}
    W(t)V_xW(t)V_x=
    \begin{cases}
        \mathbb{1}, & \text{if } [W(t),V_x]=0,\\
        -\mathbb{1}, & \text{if } \{W(t),V_x\}=0.
    \end{cases}
\end{equation}
Taking the normalized trace then gives
\begin{equation}\nonumber
    \frac{1}{2^N}\mathrm{Tr}\!\left[W(t)V_xW(t)V_x\right]
    =
    \begin{cases}
        +1, & \text{if } [W(t),V_x]=0,\\
        -1, & \text{if } \{W(t),V_x\}=0,
    \end{cases}
\end{equation}
and therefore
\begin{equation}
    C(x,t)=
    \begin{cases}
        0, & \text{if } W(t)\text{ and }V_x\text{ commute},\\
        1, & \text{if } W(t)\text{ and }V_x\text{ anticommute}.
    \end{cases}
\end{equation}

Thus, in the present setting the OTOC reduces to a sharp indicator of whether the evolved operator $W(t)$ has developed nontrivial Pauli support at site $x$.

\subsection{Interpretation in the present graph-state circuit model}

In Clifford circuits, Pauli operators evolve into Pauli strings under conjugation,
\begin{equation}
    \mathscr{U}_t^\dagger P\,\mathscr{U}_t = \pm P',
\end{equation}
where $P'$ is another Pauli string; see, e.g., Refs.~\cite{Aaronson-Gottesman}. This property underlies the efficient classical simulation of Clifford dynamics and makes the OTOC particularly simple to evaluate in the present model.

In our graph-state blocks, we only use the gates Hadamard and controlled-$Z$. Their action on local Pauli operators is
\begin{align} \nonumber
    H X H &= Z, &
    H Z H &= X,\\\nonumber
    \mathrm{CZ}_{ij}\, X_i\, \mathrm{CZ}_{ij} &= X_i Z_j, &
    \mathrm{CZ}_{ij}\, X_j\, \mathrm{CZ}_{ij} &= Z_i X_j,\\\nonumber
    \mathrm{CZ}_{ij}\, Z_i\, \mathrm{CZ}_{ij} &= Z_i, &
    \mathrm{CZ}_{ij}\, Z_j\, \mathrm{CZ}_{ij} &= Z_j.
\end{align}
These relations show that, starting from a local $X$ operator, Clifford evolution generated by the graph-state preparation circuit maps it into a Pauli string whose local components are drawn from $\{\mathbb{1},X,Z\}$, while its spatial support can spread through the circuit.

For this reason, in the main text we choose
\begin{equation}
    W = X_{N/2},
    \qquad
    V_x = Y_x.
\end{equation}
The probe operator $Y_x$ anticommutes with both $X_x$ and $Z_x$, while commuting with $\mathbb{1}_x$. Therefore, in the present circuit architecture, the OTOC $C(x,t)$ provides a direct indicator of whether the Heisenberg-evolved operator $W(t)$ has reached site $x$ with nontrivial local Pauli support.

\section{Dependence on sparsity}\label{appendix:sparsity}

\begin{figure}[t!]
    \centering
    \includegraphics[width=0.8\linewidth]{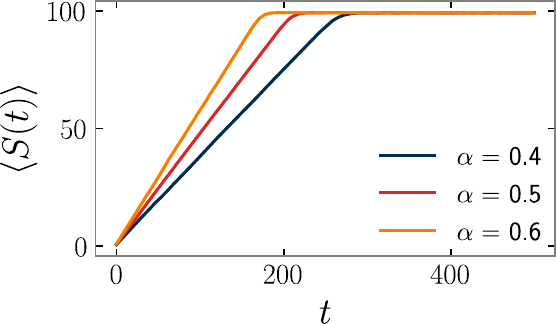}
   \caption{
    Bipartite entanglement growth $\langle S_A(t) \rangle$ for random Clifford circuits constructed from a five-qubit graph-state block (graph 1, star/GHZ-type) for different values of the sparsity parameter $\alpha$. As $\alpha$ increases (i.e., as more blocks are applied per layer), the linear growth rate of entanglement increases. System size $N=200$ and results are averaged over independent circuit realizations.
    }
    \label{fig:ent_vary_alpha}
\end{figure}

\begin{figure}[t!]
    \centering
    \includegraphics[width=0.75\linewidth]{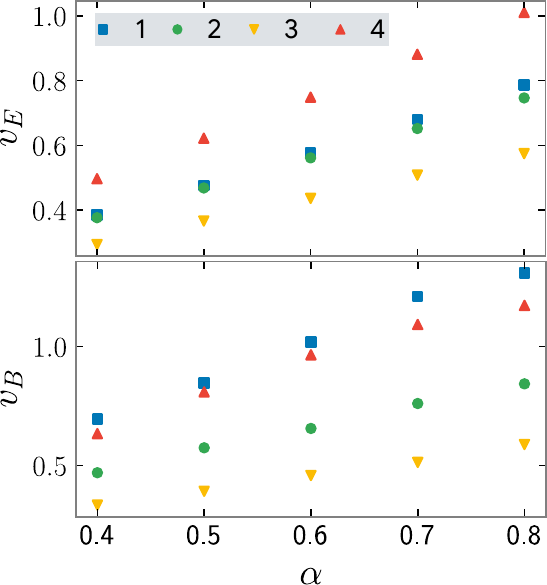}
    \caption{Entanglement velocity $v_E$ and butterfly velocity $v_B$ as functions of the sparsity parameter $\alpha$ for five-qubit graph-state blocks. Each curve corresponds to a different LC-inequivalent graph state (same labeling as in Fig.~\ref{fig:entanglement-g5&6}(a)). Both velocities increase monotonically with $\alpha$, while the relative ordering across graph states remains unchanged.
    }
   \label{fig:vel_vary_alpha}
\end{figure}

The sparsity parameter $\alpha$ controls the fraction of qubits acted on by graph-state blocks in each circuit layer. In the main text we fix $\alpha=1/2$. Here we analyze how varying $\alpha$ affects entanglement growth and operator spreading.

We begin by considering a representative five-qubit graph-state block (graph 1, corresponding to a star/GHZ-type structure) and compare different values of $\alpha$. Figure~\ref{fig:ent_vary_alpha} shows the resulting bipartite entanglement entropy $S_A(t)$ for several values of $\alpha$. As $\alpha$ increases, the slope of the linear growth regime increases, indicating faster entanglement generation.

To quantify this behavior, Fig.~\ref{fig:vel_vary_alpha} shows the extracted entanglement velocity $v_E$ and butterfly velocity $v_B$ as functions of $\alpha$ for all five-qubit graph-state blocks analyzed in the main text (see Fig.~\ref{fig:entanglement-g5&6}~(a)). Both velocities increase monotonically with $\alpha$, reflecting the higher density of local circuit updates.

Importantly, the relative ordering of graph-state blocks remains unchanged across all values of $\alpha$ considered: blocks that generate faster entanglement or operator spreading at a given sparsity continue to do so at other values of $\alpha$. Increasing $\alpha$ therefore accelerates the dynamics without altering the hierarchy induced by the internal structure of the graph-state blocks.

\section{Boundary effects}\label{appendix:obc}

\begin{figure}[t!]
    \centering
    \includegraphics[width=0.9\linewidth]{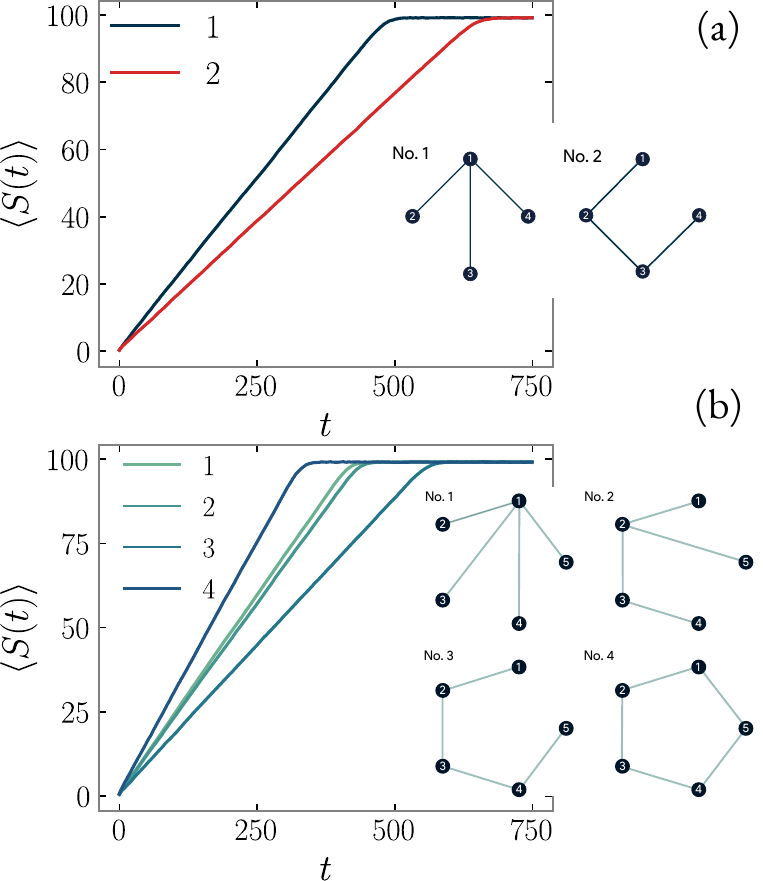}
    \caption{Bipartite entanglement growth for random Clifford circuits with open boundary conditions (OBC). 
The half-chain von Neumann entropy $S_A(t)$ is shown as a function of circuit depth $t$ for circuits built from LC–inequivalent graph-state blocks with sizes (a) $n=4$ and (b) $n=5$. 
Each curve corresponds to a different graph-state block, while the system size and sparsity are the same as in the main text. 
The qualitative behavior remains the same as in the periodic boundary condition (PBC) case: entanglement grows linearly before saturating, with slopes defining block-dependent entanglement velocities $v_E$.}
\label{fig:ent_growth_obc}
\end{figure}

To verify that the observed hierarchy of entanglement velocities is not an artifact of periodic boundary conditions, we repeat the simulations using open boundary conditions (OBC). 
Figure~\ref{fig:ent_growth_obc} shows representative entanglement growth curves for block sizes $n=4$ and $n=5$. 
The qualitative behavior remains unchanged: the entropy grows linearly before saturating, and different graph-state blocks produce distinct slopes corresponding to different entanglement velocities $v_E$. 
This indicates that the block-dependent hierarchy observed in the main text is robust against boundary effects.

\vfill

\bibliographystyle{apsrev4-1}
\bibliography{biblio}

\end{document}